\documentclass[aps,pre,reprint, amsmath, amssymb,superscriptaddress]{revtex4-1}

\usepackage{morefloats,ulem}
\usepackage{bm}
\newcommand{\beq}{\begin{equation}}
\newcommand{\eeq}{\end{equation}}

\usepackage[retainorgcmds]{IEEEtrantools}
\usepackage{graphicx,tikz,placeins}
\usepackage{mathrsfs}
\usepackage{amsmath,amssymb,amsfonts,physics}
\usepackage{color}
\usepackage{float}
\usepackage{times,txfonts}
\usepackage{nicefrac}
\usepackage[colorlinks=true,linkcolor=blue,urlcolor=blue,citecolor=blue,pdfusetitle]{hyperref}
\usepackage{physics}
\usepackage{soul}

\usepackage{ulem}
\newcommand{\maxp}{\textrm{\tiny{MP}}}
\newcommand{\maxe}{\textrm{\tiny{ME}}}

\begin{document}
\title{Powerful ordered collective heat engines}
\author{Fernando S. Filho}
\affiliation{Universidade de São Paulo,
Instituto de Física,
Rua do Matão, 1371, 05508-090
São Paulo, SP, Brazil}
\author{Gustavo A. L. For\~ao}
\affiliation{Universidade de São Paulo,
Instituto de Física,
Rua do Matão, 1371, 05508-090
São Paulo, SP, Brazil}
\author{Daniel M. Busiello}
\affiliation{ Max Planck Institute for the Physics of Complex Systems, 01187 Dresden, Germany}
\author{B. Cleuren}
\affiliation{UHasselt, Faculty of Sciences, Theory Lab, Agoralaan, 3590 Diepenbeek, Belgium}
\date{\today}
\author{Carlos E. Fiore}
\affiliation{Universidade de São Paulo,
Instituto de Física,
Rua do Matão, 1371, 05508-090
São Paulo, SP, Brazil}
\date{\today}

\begin{abstract} 
We introduce a class of stochastic engines in which the regime of units operating synchronously can boost the performance. Our approach encompasses a minimal setup composed of $N$ interacting units placed in contact with two thermal baths and subjected to a constant driving worksource. The interplay between unit synchronization and interaction leads to an efficiency at maximum power between the Carnot, $\eta_{c}$, and the Curzon-Ahlborn bound, $\eta_{CA}$. Moreover, these limits can be respectively saturated maximizing the efficiency, and by simultaneous optimization of power and efficiency.
We show that the interplay between Ising-like interactions and a collective ordered regime is crucial to operate as a heat engine.
The main system features are investigated by means of a linear analysis near equilibrium, and developing an effective discrete-state model that captures the effects of the synchronous phase. The present framework paves the way for the building of promising nonequilibrium thermal machines based on ordered structures.
\end{abstract}

\maketitle

{\it Introduction.--}The ambition to build efficient engines is not only prominent, but also pressing in thermodynamics since the pioneering work by Sadi Carnot \cite{carnot1978reflexions}, and gained new momentum with the development of non-equilibrium thermodynamics of small-scale systems \cite{curzon1975efficiency,seifert2012stochastic}. Unlike  thermodynamics, fluctuations become fundamental at the nano-scale and the study of their role attracted large attention, both theoretically \cite{gallavotti1995dynamical,kurchan1998fluctuation,jarzynski1997nonequilibrium,crooks1999entropy} and experimentally \cite{collin2005verification,PhysRevLett.120.010601,PhysRevLett.128.050603}. As irreversibility is unavoidable, the search for new strategies in the realm of nonequilibrium stochastic thermodynamics is crucial and strongly desirable.
Bearing this in mind, several distinct approaches have
been proposed. Among them, we highlight the study of the
maximum attainable power \cite{verley2014unlikely, 
cleuren2015universality, van2005thermodynamic, 
seifert2011efficiency, 
golubeva2012efficiency, 
karel2016, 
ciliberto2017experiments,bonanca2019,campisi2016power} and efficiency \cite{karel2016prl,mamede2021obtaining}, the modulation of the system-bath interaction time \cite{noa2021efficient,harunari2020maximal}, 
and the dynamical control via shortcuts to adiabaticy \cite{RevModPhys.91.045001,deffner2020thermodynamic,pancotti2020speed} or isothermality \cite{2206.02337}. 
 
The above examples deal with engines composed of a single or a few units. However, nature is plenty of complex systems composed of many interacting entities, in which cooperative effects often play a crucial role. Examples span multiple biological scales \cite{gnesotto2018broken}, from microbes \cite{smith2019public} to the human brain \cite{lynn2021broken}, and have been studied in a broad range of research fields, from non-equilibrium effects in chemical processes \cite{rao2016nonequilibrium,busiello2021dissipation,dass2021equilibrium} to synchronization in biological networks \cite{bonifazi2009gabaergic,schneidman2006weak,buzsaki2014log,gal2017rich,tonjes2021coherence}.
This vast spectrum of applications highlights that the demand for implementable and robust optimal strategies to engineer collective engines is important and timely.  Although the interplay between collective effects and system's performances has been extensively studied in quantum systems \cite{mukherjee2021manybody,niedenzu2018cooperative,PhysRevApplied.19.034023,latune2020collective,kamimura2022collective,mavocei2022performance},
the development of classical setups built from interacting units is comparatively much less known and still remains at a primary stage \cite{gatien,herpich,herpich2,campisi2016power,sune2019out}.

In this letter, we introduce a general class of collective engines, inspired by ferromagnetic equilibrium models \cite{yeomans1992statistical,RevModPhys.54.235,PhysRevA.4.1071,PhysRevLett.67.1027}. They have  a long-standing importance in the context of collective effects and are at the heart of numerous theoretical and experimental advances, having distinct models (e.g. the Ising, Potts, XY and Heisenberg) as ideal platforms for describing ferromagnetism. Optimizing power and efficiency by changing driving and coupling parameters, we show that synchronized operations under ordered (ferromagnetic)  arrangements play a central role in improving system performances. 
The main features and optimization routes of the engine proposed here can be unveiled both using a linear analysis close to equilibrium and an effective discrete-state model capturing all relevant effects.
\begin{figure*}[t]
\centering
\includegraphics[scale=0.23]{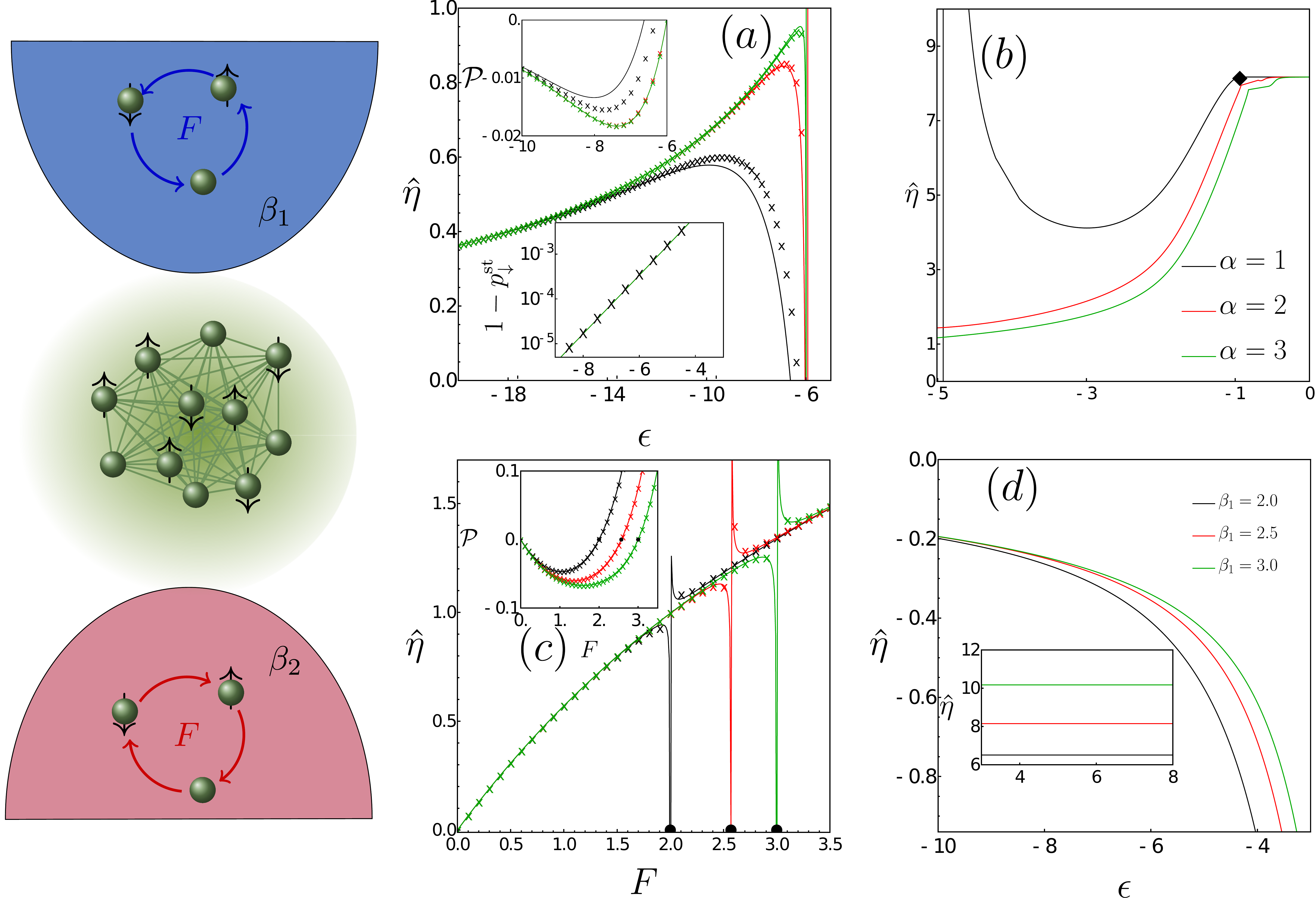} 
\caption{Left: Schematics of $q=3$ engines. Arrows in the reservoirs indicate the direction of the driving $F$, which is clockwise at high temperature and counter-clockwise at low temperature. (a) Model A ($\epsilon_{\downarrow\uparrow} \neq 0$). The efficiency ${\hat\eta}=\eta/\eta_c$ is shown for different $\alpha$ as a function of the coupling strength $\epsilon$ in the strong collective phase (smaller $\epsilon$). Lines are exact results, while dots represents the effective model. Power output per unit, ${\cal P}$, is presented in the upper inset, while the lower inset is a semilog-plot of $1-p^{\rm st}_{\downarrow}$ to show the robustness of the effective description. (b) Same as (a), but in the presence of weak collective effects (larger $\epsilon$). The symbol $\blacklozenge$ in (b) indicates the critical point $\epsilon_c$ separating the regimes of collective and independent units. As a result, collective ordered operations favor a heat engine behavior. Parameters in (a) and (b): $\beta_1=2$, $\beta_2=1$ and $F=2$. (c) Model A. ${\hat \eta}$ and ${\cal P}$ (inset) versus $F$ for different $\beta_1$. Vertical lines mark the crossover between heat engine and pump regimes, also indicated by $\bullet$. Parameters in (c): $\alpha=3$,$\epsilon=-6,\beta_2=1$. As previously, symbols correspond to the effective model. (d) Model B ($\epsilon_{\downarrow\uparrow} = 0$). For $F=1$, $\beta_2=1$ and different $\beta_1$, the efficiency is shown as a function of $\epsilon$, indicating only a dud regime in this case, as Ising-like interactions are absent. As $\epsilon$ increases, model B shows a pump behavior (inset).}
\label{fig1}
\end{figure*}

{\it Thermodynamics.--}
Since our goal is to investigate main features and advantages of the cooperative behavior emerging from ordered agents, we design a system composed of $N$ all-to-all interacting units. Each unit can occupy $q$ different states, so that a microstate $i$ of the system is an $N$-dimensional vector containing the states of all units. 
This system is placed in contact with two baths at different temperatures ($\nu = 1$ is the cold one, $\nu = 2$ the hot) to work as a heat engine. Moreover, worksources originate from $\gamma_F$ distinct driving forces that also depend on the bath, i.e., $F_\gamma^{(\nu)}$ with $\gamma = 1, \dots, \gamma_F$. In Fig.~\ref{fig1}a, we present a sketch of the model for $q=3$.
The dynamics of microstates is governed by the master equation:
\begin{eqnarray}
\dot p_j=\sum_{\nu=1}^2\sum_{i\neq j}(\omega^{(\nu)}_{ji}p_i-\omega^{(\nu)}_{ij}p_j), \quad \omega^{(\nu)}_{ji}=\Gamma e^{-\frac{\beta_\nu}{2}\{E_j-E_i+\sum_\gamma F_\gamma^{(\nu)}d_{\gamma,ji}^{(\nu)}\}} \nonumber
\end{eqnarray}
where $\omega^{(\nu)}_{ji}$ is the transition rate from $i$ to $j$ due to the bath $\nu$, and $d^{(\nu)}_{\gamma,ij}$ are anti-symmetric coefficients associated with non-conservative driving. Denoting by $N^{(i)}_\beta$, $\beta = 1, \dots, q$, the occupation number of the state $\beta$ in the microstate $i$, a transition to $j$ leads to $N^{(j)}_\beta = N^{(i)}_\beta - 1$ and $N^{(j)}_{\beta'} = N^{(i)}_{\beta'} + 1$, where $\beta$ and $\beta'$ depend on initial and final microstates. Clearly, to map microstates into occupation numbers, we need to perform a coarse-graining procedure (see Supplemental Material). The total energy $E_i$ is given by the all-to-all expression:
\begin{equation}
E_{i}=\sum_{\beta=1}^q\epsilon_{\beta} N^{(i)}_{\beta}+\frac{1}{2N}\sum_{(\beta,\beta'<\beta)}^q\Big[\epsilon_{\beta \beta}N^{(i)}_{\beta}(N^{(i)}_{\beta}-1)+ 2\epsilon_{\beta\beta'}N^{(i)}_{\beta}N_{\beta'}   \Big],
\label{gen}
\end{equation}
where $\epsilon_\beta,\epsilon_{\beta\beta}$ and $\epsilon_{\beta\beta'}$ denote individual and interaction energies for units in the same and different states, respectively.
From these preliminaries, the first law of thermodynamics is formulated from the time evolution of mean energy $\langle E\rangle=\sum_{j}E_j p_j$, which is given by $d\langle {E}\rangle/dt=\langle {\mathtt P}\rangle +\langle \dot{{\mathtt Q}}_1\rangle+\langle\dot{{\mathtt Q}}_2\rangle$, where the mean power $\langle {\mathtt P}\rangle$ and heat fluxes from the bath $\nu$ are
\begin{eqnarray}
\label{work}
\langle {\mathtt P}\rangle &=& -\sum_{(\nu,\gamma)} F_\gamma^{(\nu)}\sum_{(i,j)}d_{\gamma,ji}^{(\nu)}J_{ji}^{(\nu)},\\
\langle \dot{{\mathtt Q}}_\nu\rangle &=& \sum_{(i,j)}\left(E_j-E_i+\sum_{\gamma}F_\gamma^{(\nu)}d_{\gamma,ji}^{(\nu)}\right)J_{ji}^{(\nu)}\;,
\label{heat}
\end{eqnarray}
expressed in terms of the probability current $J_{ij}^{(\nu)} =\omega^{(\nu)}_{ji}p_i-\omega^{(\nu)}_{ij}p_j$. The nonequilibrium steady state (NESS) is characterized by the probabilities $\{p^{\rm st}_j\}$ satisfying $ \langle {\mathtt P}\rangle+\langle\dot{{\mathtt Q}}_1\rangle+\langle\dot{\mathtt Q}_2\rangle=0$ and associated with a positive entropy production into the environment 
$\langle {\dot \sigma} \rangle=-\beta_1\langle\dot{{\mathtt Q}}_1\rangle-\beta_2\langle\dot{{\mathtt Q}}_2\rangle$.
Although exact, $\langle {\dot \sigma} \rangle $ can be further simplified when some channels are faster than others \cite{busiello2020coarse}.
Employing the steady-state condition, $\langle {\dot \sigma} \rangle$ can be rewritten as $\langle {\dot \sigma} \rangle = \beta_1\langle {\mathtt P}\rangle+\left(\beta_1 - \beta_2\right) \langle \dot{{\mathtt Q}}_2\rangle$,  allowing us to characterize the engine performance through two (equivalent) definitions of efficiency, $\eta=-\langle{\mathtt P}\rangle/\langle\dot{{\mathtt Q}}_2\rangle$ and  from the entropy production, ${\hat \eta} = -\eta_{\textrm{c}}^{-1}\langle{\mathtt P}\rangle /\langle \dot{{\mathtt Q}}_2\rangle$, solely differing from each other for the Carnot bound $\eta_{\textrm{c}}=1-\beta_2/\beta_1$. 
A heat engine partially converts the heat extracted from the hot thermal bath ($\langle\dot{{\mathtt Q}}_2\rangle>0$) into power output ($\langle{\mathtt P}\rangle<0$), hence
exhibiting, by construction, a positive and bounded efficiency, $0\le \eta \le \eta_c$ ($0\le {\hat \eta}\le 1$). Conversely, the pump regime is characterized by an amount of work $\langle{\mathtt P}\rangle>0$ which is partially used to sustain a heat flux from the cold to hot bath, i.e., $\langle\dot{{\mathtt Q}}_2 \rangle<0$, hence 
$\eta_c<\eta<\infty$ ($1<{\hat \eta}<\infty$). Finally, for $\eta<0$ the engine works in the so-called dud regime, i.e., the engine does not generate power.
The analysis will be first carried out for $N\rightarrow \infty$, deriving the evolution for the mean occupation density of the state $\beta$, $p_\beta = \langle \sum_i N^{(i)}_\beta/N \rangle$ and next for finite $N$, studying how finite-size effects disappears to converge to a mean field behavior.

{\it Minimal ordered collective heat engines and the effective
description.--} Eq.~\eqref{gen} presents a huge number of parameters to be considered, precisely $2q+q(q-1)/2$. For simplicity, we restrict our analysis to the cases
$q=2$ and $q=3$, which can be respectively mapped into spin models $S=1/2$, $\beta = \{\downarrow,\uparrow\}$, and $S=1$, $\beta = \{\downarrow,0,\uparrow\}$. We also consider two choices for the interaction parameters $\epsilon_{\beta\beta'}$, inspired by two cornerstones in statistical physics, Ising and Potts models \cite{yeomans1992statistical,salinas2001introduction}, here respectively named model A and B, for simplicity. Hence, in model A with $q=3$, we take $\epsilon_{\uparrow\uparrow}=\epsilon_{\downarrow\downarrow}=\epsilon$, $\epsilon_{\uparrow\downarrow}=-\alpha\epsilon$, and $\epsilon_{\uparrow 0} = \epsilon_{\downarrow 0} = \epsilon_{0 0} = 0$, with $\epsilon_{\beta\beta'}$ symmetric for every $\beta$, $\beta'$. Here, $\alpha$ tunes the interaction strength between units in different states. Conversely, model B is defined by $\epsilon_{\beta\beta'}=\epsilon\delta_{\beta,\beta'}$, with no interaction between units in different states. We always consider the self-interaction terms, $\epsilon_\beta$, to be all equal. In analogy to other engine setups \cite{gatien,herpich}, and also compatibly with models of biochemical motors, such as kinesin \cite{liepelt1,liepelt2}, photo-acids \cite{berton2020thermodynamics}, and ATP-driven chaperones \cite{de2014hsp70}, the worksource is implemented by introducing a bias for the occurrence of certain transitions, forcing, in this context, each unit to rotate in its state-space (see Fig.~\ref{fig1} for $q=3$ and Supplemental Material). Practically, this bias is realized by setting $d_{\gamma,ij}^{(\nu)}=(-1)^\nu$ if the transition from $j$ to $i$ is clockwise, where the opposite rate is determined by the anti-symmetric property. We further simplify the system by taking $\Gamma = 1$ and only one kind of driving, i.e., $\gamma_F=1$ and $F^{(\nu)}=F$ for both $\nu$.

Fig.~\ref{fig1} shows the main features of model A and B for $q=3$ and $N\rightarrow \infty$, in which $p^{\rm st}_\beta \in \{ p^{\rm st}_\uparrow, p^{\rm st}_\downarrow, p^{\rm st}_0 \}$. In such a mean field limit, the system is described by a non-linear master equation that cannot be self-consistently solved (see Supplemental Material). In Fig.~\ref{fig1}(a)-(b), we show efficiency and power output per unit for model A. This setting allows for the existence of a collective ordered phase for large negative $\epsilon$. In this regime, the system behaves as a heat engine Fig.~\ref{fig1}(a). As $\epsilon$ increases, units deviate from a synchronized phase and a pump behavior emerges. For $\alpha=1$, units starts operating independently after a phase transition $\blacklozenge$ in Fig.~\ref{fig1}(b), while for other $\alpha$ there is a crossover between these collective and independent regimes. Moreover, Fig.~\ref{fig1}(c) shows that $F$ can be used as a parameter to control the system, as when $F$ increases a pump behavior emerges even in the collective ordered phase. As shown in Supplemental material, power and heat fluxes are independent from $\epsilon$ when units operate independently, indicating that, in the collective phase, $\epsilon$ can be chosen appropriately to lead to a better performance even as a pump and hence hinting at the relevance of a synchronous phase for this class of engines.  Conversely, no heat regime is present for model B Fig. \ref{fig1}(d), when only Potts-like interactions are present. If units operate independently, the engine can only work as a pump in this case. Although model B has been proposed as a work-to-work converter \cite{herpich}, the absence of Ising-like interactions makes the synchronous phase useless to operate as a heat engine. For these reasons, model A will be used as reference model from now on. Analogous findings are also reported in the Supplemental Material for $q=2$.

Despite exact in all case, the non-linear form of the master equation prevents the derivation of closed expressions for probabilities and clear insights about the influence of each parameter. To grasp the main features of the system in the regime of strong collective effects, we develop an effective discrete-state description which is valid in the ordered collective regime. By taking $\alpha > 0$, $\epsilon < 0$, and $- \epsilon \gg F > 0$, this effective description can be derived  employing the matrix-tree theorem \cite{schnakenberg}. We start describing the system as a coarse-grained $q$-state model. Considering $q=2$ just to fix the ideas, we expand the transition rates up to the leading order around $p^{st}_{\downarrow} \approx 1$ and $p^{st}_{\uparrow} \approx 0$, or vice-versa, to obtain the desired approximated results. For $q=2$, the probability $p^{\rm st}_\downarrow$ is approximately given by $p^{st}_{\downarrow}\approx 1-e^{\frac{1}{2}\{(\beta_1+\beta_2)(1+\alpha)\epsilon+(\beta_1-\beta_2)F\}}$, with the system synchronization given by $|M| = 2p^{\rm st}_\downarrow -1$ and close to $1$. The main features and expressions for $q=2$ are derived in Supplemental Material.
\begin{figure}[t]
\centering
\includegraphics[scale=0.25]{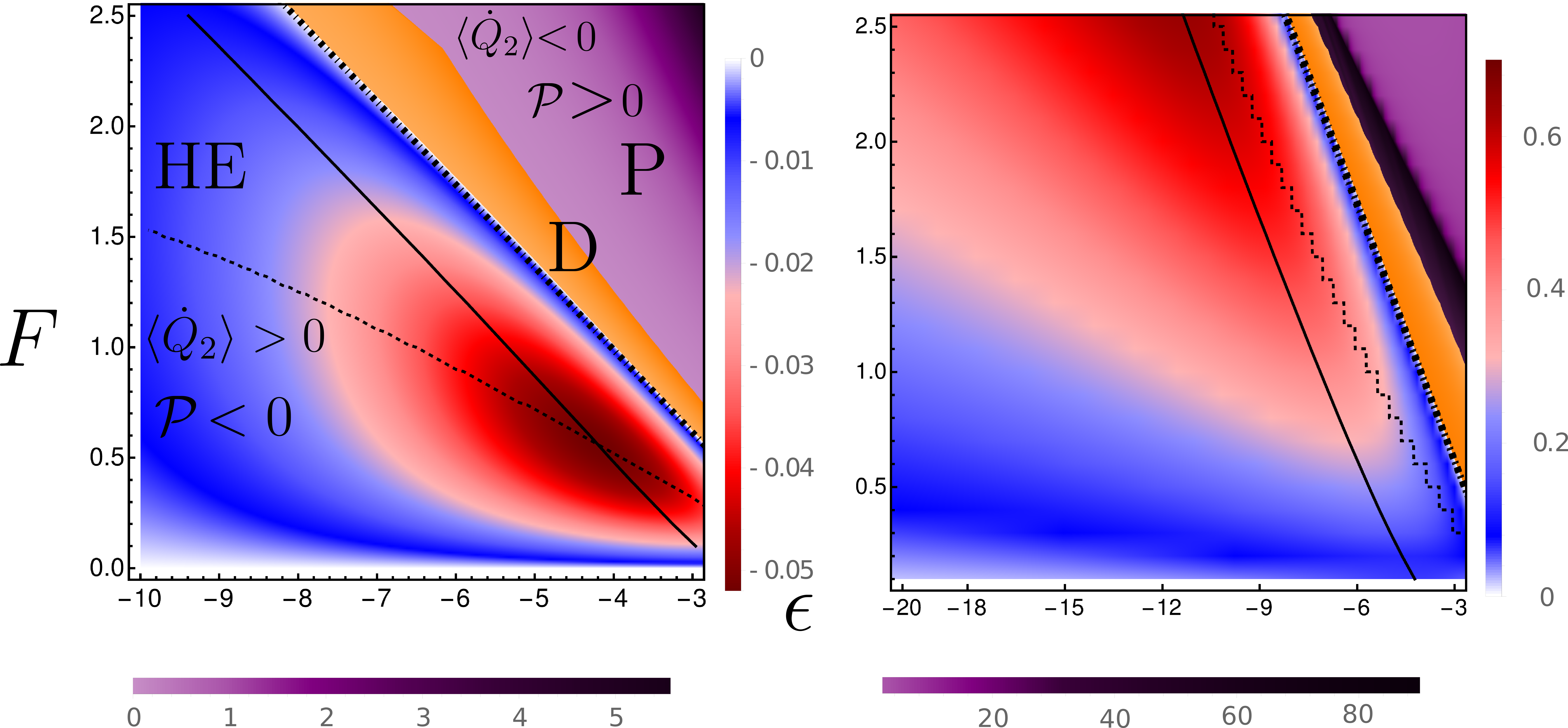}
\caption{Model A and $q=3$. Left panel depicts the power heat map as a function of driving $F$ and coupling $\epsilon$. HE, P and D indicate, respectively, heat engine, pump, and dud regimes. The solid line shows the maximum power with respect to $F$ at fixed $\epsilon$, while the dashed line accounts for the maximization with respect to $\epsilon$. These two lines cross at the global maximum power. Right panel shows the efficiency $\hat{\eta}$ heat map as a function of $F$ and $\epsilon$. Solid and dashed lines again indicate maximization with respect to $F$ and $\epsilon$, respectively. In both panels, the dot-dashed lines only indicate the boundaries of heat engine regimes. Parameters: $\beta_1=2, \beta_2=1, \alpha=1$.}
\label{fig2}
\end{figure}
Our effective description can be employed also when $q=3$, always considering model A. In this case, we followed the same steps as in the $q=2$ scenario, noting that $p^{st}_\uparrow \approx 0$. Hence, by considering $\uparrow$ as a source when $|\epsilon| \to \infty$ and $-\epsilon \gg F$, and employing a pseudo-equilibrium approximation (see Supplemental Material), one obtains that
$p^{st}_\downarrow \approx 1-e^{\frac{1}{2}\{(\beta_1+\beta_2)\epsilon+(\beta_1-\beta_2)F\}}$. The corresponding expression for power per unit is given by
\begin{eqnarray}
{\cal P}_{\rm eff} &=& F \Big[(1+M) \left(e^{\frac{\beta_1}{2} \Phi^{(\alpha)}_-}-e^{-\frac{\beta_2}{2} \Phi^{(\alpha)}_+ }-e^{-\frac{\beta_1}{2} \Phi_-}+e^{\frac{\beta_2}{2} \Phi_+}\right) \\
- &M& \left(e^{\frac{\beta_1}{2} \Phi_-}-e^{-\frac{\beta_2}{2} \Phi_+}-e^{-\frac{\beta_1}{2} (\Phi^{(\alpha)}_+ + M\epsilon)}+e^{\frac{\beta_2}{2} (\Phi^{(\alpha)}_- - M\epsilon)}\right)\Big] \nonumber \;,
\label{apow}
\end{eqnarray}
\normalsize
where $\Phi_\pm = F \pm M\epsilon$ and $\Phi^{(\alpha)}_\pm = F \pm \alpha M \epsilon$, with $M = p^{st}_\downarrow-p^{st}_\uparrow \approx 1-e^{\frac{1}{2}\{(\beta_1+\beta_2)\epsilon+(\beta_1-\beta_2)F\}}>0$. Also in this case, $\langle Q_2\rangle_{\rm eff} $ is shown in Supplemental Material. The efficiency is readily evaluated taking their ratio. In the limit of large $\alpha$, it reads:
\begin{equation}
\eta_{\rm eff}=\frac{F}{\Phi_+}\left[1-\frac{\exp\left(-\frac{\beta_1}{2}\Phi_-\right)+2M\cosh\left(\frac{\beta_1}{2}\Phi_-\right)}{\exp\left(\frac{\beta_2}{2}\Phi_+\right)+2M\cosh\left(\frac{\beta_2}{2}\Phi_+\right)}\right].
\label{effl}
\end{equation}

The validity of this approach is shown in Fig.~\ref{fig1} for different values of $\alpha$ (symbols). The effective discrete-state model provides a very good description of both the heat engine and pump regimes. However, small discrepancies arise when $p^{st}_\uparrow$ is not negligible (e.g. for small $-\epsilon$ and $F$).

The main features of the system proposed here can also be investigated through a linear analysis close to equilibrium, e.g., $\beta_1-\beta_2\ll 1$ and $F \ll 1$. In this scenario, the entropy production acquires a bilinear form and can be expressed in terms of Onsager coefficients. As described in Supplemental Material, the maximum efficiency, $\eta_{ME}$, and the efficiency at maximum power, $\eta_{MP}$, solely depend on the coupling parameter $\kappa=L_{12}/\sqrt{L_{11}L_{22}}$. Since $|\kappa|$ monotonically increases with $\epsilon$, both $\eta_{ME}$ and $\eta_{MP}$ approach to their ideal values when collective ordered effects are stronger, highlighting the importance of unit synchronization to increase engine performance.

We extend our results to a wider spectrum of values of the coupling parameter $\epsilon$ and the driving $F$. Fig.~\ref{fig2} shows the resulting heat map again for $q=3$ and model A. Heat engine (blue-red) and pump (purple) regimes are separated by an intermediate region in which units operate dudly (orange). Power and efficiency can be optimized with respect to $F$ (solid lines) or $\epsilon$ (dashed lines), where the other quantity is held fixed. It is worth noting that the power output in the heat engine regime presents a global maximum where the two optimization lines cross (dark red spot). This point coincides with the power obtained by simultaneous optimization with respect to $F$ and $\epsilon$. Conversely, no optimal point exists for the efficiency in the $(F, \epsilon)$ space, and the heat engine operates more efficiently as $|\epsilon|$ and $F$ are increased. This result hints at the possibility to boost the performance of a stochastic heat engine by favoring the emergence of collective order.

\begin{figure}[t]
\centering
\includegraphics[scale=0.42]{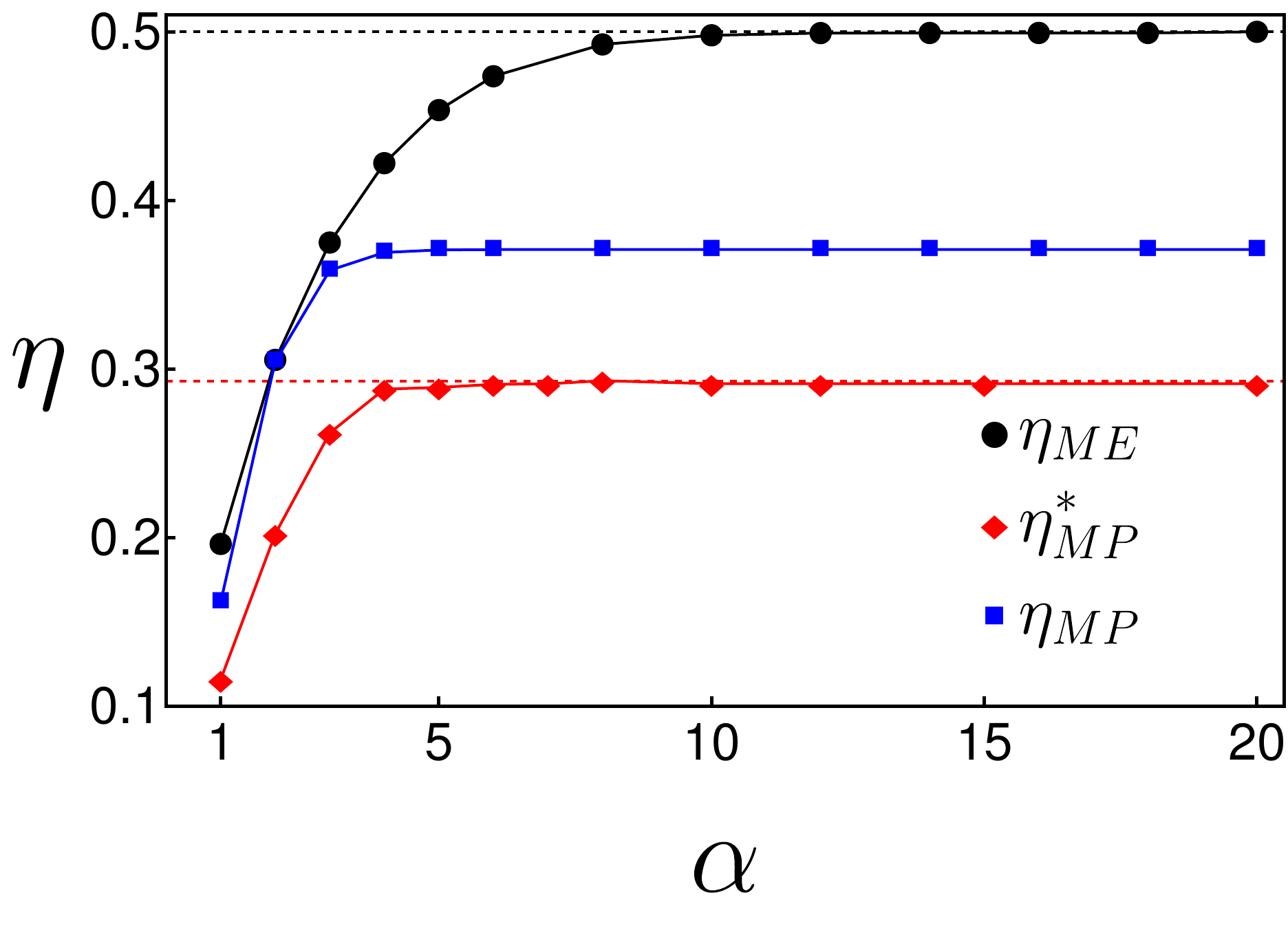} 
\caption{Maximum efficiency $\eta_{ME}$ (black circles), efficiency at maximum power $\eta_{MP}$ (blue squares) and efficiency at global maximum power $\eta^{*}_{MP}$ (red diamonds) as a function of the coupling between different states, $\alpha$. Solid lines are guides for the eye. The black and red dashed lines correspond Carnot $\eta_c$ and Curzon-Ahlborn $\eta_{CA}$ efficiencies, respectively.}
\label{effvsalpha}
\end{figure}

As suggested by Fig.~\ref{fig1}, an alternative route for optimization prescribes, at finite $\epsilon$ and $F$, to increase the value of the coupling between different states, $\alpha$. In Fig.~\ref{effvsalpha}, we show the maximum efficiency, $\eta_{ME}$, the efficiency at maximum power, $\eta_{MP}$, and the one obtained by simultaneous optimization, $\eta^*_{MP}$, as a function of $\alpha$. It is worth noting that $\eta_{ME}$ approaches (and eventually reaches) the ideal Carnot efficiency $\eta_c$, while $\eta^*_{MP}$ saturates the Curzon-Ahlborn bound, $\eta_{CA} = 1-\sqrt{\beta_2/\beta_1}$, as the coupling strength is increased. Furthermore, the efficiency at maximum power lies between these two bounds: $\eta_{CA} < \eta_{MP} < \eta_{c}$. Together with the results in Fig.~\ref{fig2}, we can state that, considering a general interacting model admitting an ordered phase, the performance as a heat engine benefits from a synchronized behavior, in combinations with the presence of Ising-like couplings.
    
{\it Many versus few interacting units and beyond the all-to-all case.--}  Although our findings have been derived in the $N \to \infty$ limit for all-to-all interactions, the main hallmarks have found to be robust when finite-size effects and other topologies are considered (see Supplemental Material). Panels (a) and (b) of Fig.~\ref{figGillespie} show, for model A and $q=3$, numerical results for the all-to-all case at increasing $N$. We notice a reduced range of parameters for which the system operates as a heat engine, but no significant qualitative changes. In the Supplemental Material, we also explore the limiting case $N=2$, finding similar results. It is worth noting that the system starts approaching the mean field behavior already for $N > 10$, in similarity to work-to-work transducers \cite{herpich}. A very interesting feature is an increasing in the finite-size efficiency, due to the fact that $\langle \dot{{\mathtt Q}}_2\rangle/N$ monotonically
decreases with $N$, while the absolute value of ${\cal P}=\langle {\mathtt P}\rangle/N$ increases in a certain range of parameters. 
The all-to-all case also describes
very precisely the behavior of interactions forming a regular arrangement. In panels $(c)$ and $(d)$ of Fig.~\ref{figGillespie} we present the case of square lattice with (increasing) $N$ units. This observation not only reinforces the generality of the model proposed here to grasp the interplay between collective effects and system's performance, but also the reliability of our results for finite-size systems.

\begin{figure}[t]
\centering
\includegraphics[scale=0.28]{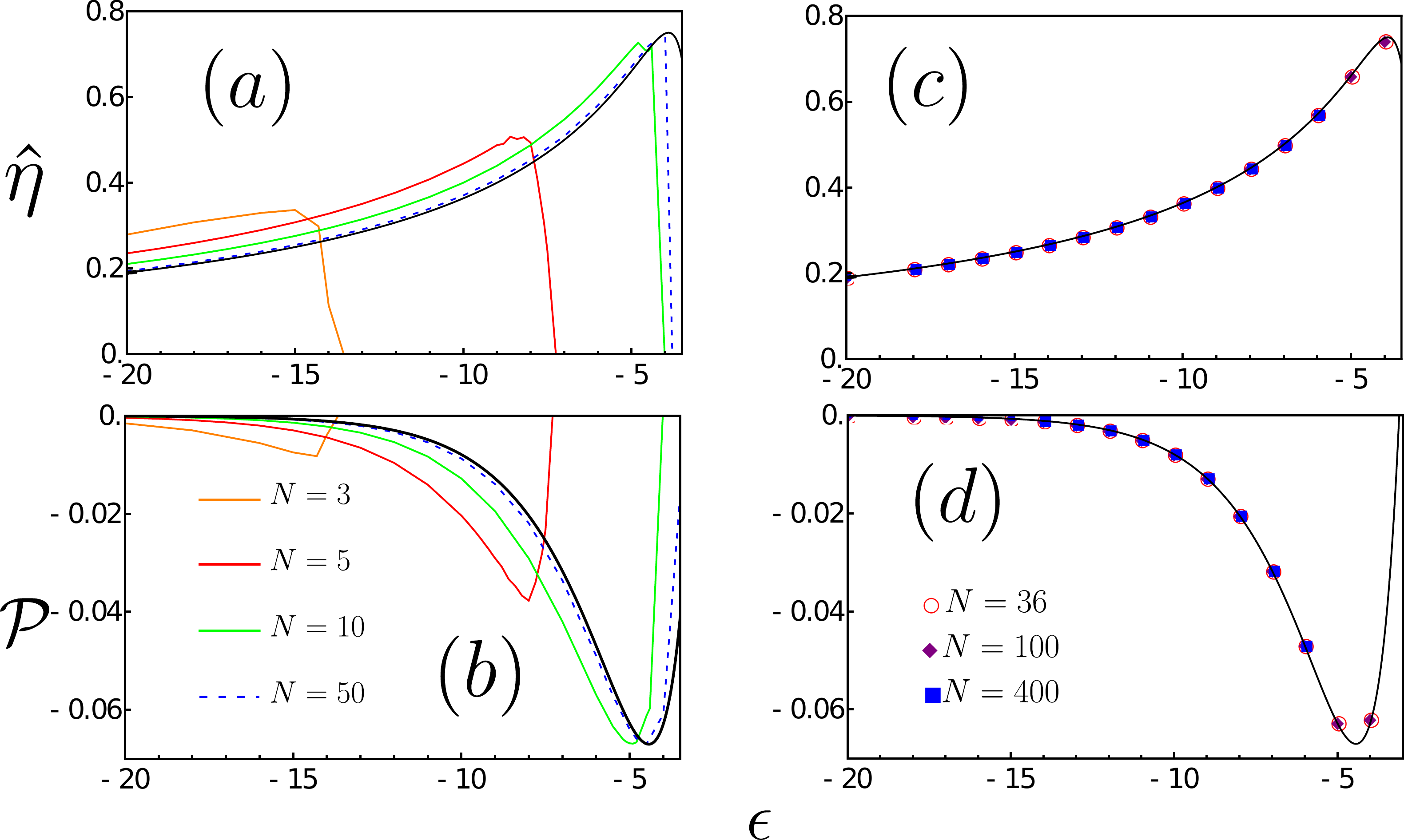} 
\caption{(a-b) Efficiency ${\hat \eta}$ and power output per unit ${\cal P}=\langle {\mathtt P}\rangle /N$ in the heat engine regime for increasing system size $N$. The black continuous line represents the $N\to\infty$ case.(c-d) The all-to-all case (continuous line) is compared with a square lattice of increasing size $N$ (dots). Numerical values have been obtained through Gillespie algorithm. Parameters: $\beta_1=2$, $\beta_2=1$, $F=1$, and $\alpha = 3$.}
\label{figGillespie}
\end{figure}

{\it Conclusions.--}We introduced a minimal class of reliable thermal engines composed of several interacting units. We showed that, when they operate in a synchronized way, the engine can exhibit distinct regimes, along with maximal powers and efficiencies, in stark contrast to what happens when units operate independently. Despite the non-trivial interplay between interactions, driving, and collective effects, all main features can be captured using linear analysis and a discrete-state effective model which proved to be very useful to characterize these engines. Our results clearly shows the importance of a collective ordered phase to have powerful stochastic heat engines. The overall approach presented here is very general and opens the door to exciting directions for future research. First, the extension to different network topologies might be important not only to build more realistic and possibly more efficient setups, but also to check the robustness of these results, obtained in the case of all-to-all interactions, when the couplings are more sparse. Furthermore, it will be interesting to draw a comparison with other stochastic engine models, such as the sequential ones, in which the system is subjected to distinct conditions for different time periods and not simultaneously. Finally, a very fascinating open question remains to set universal bounds for power, efficiency and dissipation, possibly expressed in terms of interaction parameters and strength of collective effects. They might provide important insights about the importance of synchronized operations to boost the performance of interacting systems in different contexts, from biochemical engines \cite{julicher1995cooperative, tu2008nonequilibrium, horowitz2014thermodynamics} to information processing \cite{nicoletti2021mutual, tkavcik2016information}.

{\it Acknowledgments.--}
Authors are grateful to Pedro Harunari for useful suggestions and comments. This work has received the financial support from CAPES and FAPESP under grants 2021/03372-2 and 2021/13287-2, respectively. We also acknowledge CNPq for financial support.

\bibliography{refs}

\begin{thebibliography}{76}%
\makeatletter
\providecommand \@ifxundefined [1]{%
 \@ifx{#1\undefined}
}%
\providecommand \@ifnum [1]{%
 \ifnum #1\expandafter \@firstoftwo
 \else \expandafter \@secondoftwo
 \fi
}%
\providecommand \@ifx [1]{%
 \ifx #1\expandafter \@firstoftwo
 \else \expandafter \@secondoftwo
 \fi
}%
\providecommand \natexlab [1]{#1}%
\providecommand \enquote  [1]{``#1''}%
\providecommand \bibnamefont  [1]{#1}%
\providecommand \bibfnamefont [1]{#1}%
\providecommand \citenamefont [1]{#1}%
\providecommand \href@noop [0]{\@secondoftwo}%
\providecommand \href [0]{\begingroup \@sanitize@url \@href}%
\providecommand \@href[1]{\@@startlink{#1}\@@href}%
\providecommand \@@href[1]{\endgroup#1\@@endlink}%
\providecommand \@sanitize@url [0]{\catcode `\\12\catcode `\$12\catcode
  `\&12\catcode `\#12\catcode `\^12\catcode `\_12\catcode `\%12\relax}%
\providecommand \@@startlink[1]{}%
\providecommand \@@endlink[0]{}%
\providecommand \url  [0]{\begingroup\@sanitize@url \@url }%
\providecommand \@url [1]{\endgroup\@href {#1}{\urlprefix }}%
\providecommand \urlprefix  [0]{URL }%
\providecommand \Eprint [0]{\href }%
\providecommand \doibase [0]{http://dx.doi.org/}%
\providecommand \selectlanguage [0]{\@gobble}%
\providecommand \bibinfo  [0]{\@secondoftwo}%
\providecommand \bibfield  [0]{\@secondoftwo}%
\providecommand \translation [1]{[#1]}%
\providecommand \BibitemOpen [0]{}%
\providecommand \bibitemStop [0]{}%
\providecommand \bibitemNoStop [0]{.\EOS\space}%
\providecommand \EOS [0]{\spacefactor3000\relax}%
\providecommand \BibitemShut  [1]{\csname bibitem#1\endcsname}%
\let\auto@bib@innerbib\@empty
\bibitem [{\citenamefont {Carnot}(1978)}]{carnot1978reflexions}%
  \BibitemOpen
  \bibfield  {author} {\bibinfo {author} {\bibfnamefont {S.}~\bibnamefont
  {Carnot}},\ }\href@noop {} {\emph {\bibinfo {title} {R{\'e}flexions sur la
  puissance motrice du feu}}},\ \bibinfo {number} {26}\ (\bibinfo  {publisher}
  {Vrin},\ \bibinfo {year} {1978})\BibitemShut {NoStop}%
\bibitem [{\citenamefont {Curzon}\ and\ \citenamefont
  {Ahlborn}(1975)}]{curzon1975efficiency}%
  \BibitemOpen
  \bibfield  {author} {\bibinfo {author} {\bibfnamefont {F.}~\bibnamefont
  {Curzon}}\ and\ \bibinfo {author} {\bibfnamefont {B.}~\bibnamefont
  {Ahlborn}},\ }\href@noop {} {\bibfield  {journal} {\bibinfo  {journal}
  {American Journal of Physics}\ }\textbf {\bibinfo {volume} {43}},\ \bibinfo
  {pages} {22} (\bibinfo {year} {1975})}\BibitemShut {NoStop}%
\bibitem [{\citenamefont {Seifert}(2012)}]{seifert2012stochastic}%
  \BibitemOpen
  \bibfield  {author} {\bibinfo {author} {\bibfnamefont {U.}~\bibnamefont
  {Seifert}},\ }\href@noop {} {\bibfield  {journal} {\bibinfo  {journal}
  {Reports on progress in physics}\ }\textbf {\bibinfo {volume} {75}},\
  \bibinfo {pages} {126001} (\bibinfo {year} {2012})}\BibitemShut {NoStop}%
\bibitem [{\citenamefont {Gallavotti}\ and\ \citenamefont
  {Cohen}(1995)}]{gallavotti1995dynamical}%
  \BibitemOpen
  \bibfield  {author} {\bibinfo {author} {\bibfnamefont {G.}~\bibnamefont
  {Gallavotti}}\ and\ \bibinfo {author} {\bibfnamefont {E.~G.~D.}\ \bibnamefont
  {Cohen}},\ }\href@noop {} {\bibfield  {journal} {\bibinfo  {journal}
  {Physical review letters}\ }\textbf {\bibinfo {volume} {74}},\ \bibinfo
  {pages} {2694} (\bibinfo {year} {1995})}\BibitemShut {NoStop}%
\bibitem [{\citenamefont {Kurchan}(1998)}]{kurchan1998fluctuation}%
  \BibitemOpen
  \bibfield  {author} {\bibinfo {author} {\bibfnamefont {J.}~\bibnamefont
  {Kurchan}},\ }\href@noop {} {\bibfield  {journal} {\bibinfo  {journal}
  {Journal of Physics A: Mathematical and General}\ }\textbf {\bibinfo {volume}
  {31}},\ \bibinfo {pages} {3719} (\bibinfo {year} {1998})}\BibitemShut
  {NoStop}%
\bibitem [{\citenamefont {Jarzynski}(1997)}]{jarzynski1997nonequilibrium}%
  \BibitemOpen
  \bibfield  {author} {\bibinfo {author} {\bibfnamefont {C.}~\bibnamefont
  {Jarzynski}},\ }\href@noop {} {\bibfield  {journal} {\bibinfo  {journal}
  {Physical Review Letters}\ }\textbf {\bibinfo {volume} {78}},\ \bibinfo
  {pages} {2690} (\bibinfo {year} {1997})}\BibitemShut {NoStop}%
\bibitem [{\citenamefont {Crooks}(1999)}]{crooks1999entropy}%
  \BibitemOpen
  \bibfield  {author} {\bibinfo {author} {\bibfnamefont {G.~E.}\ \bibnamefont
  {Crooks}},\ }\href@noop {} {\bibfield  {journal} {\bibinfo  {journal}
  {Physical Review E}\ }\textbf {\bibinfo {volume} {60}},\ \bibinfo {pages}
  {2721} (\bibinfo {year} {1999})}\BibitemShut {NoStop}%
\bibitem [{\citenamefont {Collin}\ \emph {et~al.}(2005)\citenamefont {Collin},
  \citenamefont {Ritort}, \citenamefont {Jarzynski}, \citenamefont {Smith},
  \citenamefont {Tinoco},\ and\ \citenamefont
  {Bustamante}}]{collin2005verification}%
  \BibitemOpen
  \bibfield  {author} {\bibinfo {author} {\bibfnamefont {D.}~\bibnamefont
  {Collin}}, \bibinfo {author} {\bibfnamefont {F.}~\bibnamefont {Ritort}},
  \bibinfo {author} {\bibfnamefont {C.}~\bibnamefont {Jarzynski}}, \bibinfo
  {author} {\bibfnamefont {S.~B.}\ \bibnamefont {Smith}}, \bibinfo {author}
  {\bibfnamefont {I.}~\bibnamefont {Tinoco}}, \ and\ \bibinfo {author}
  {\bibfnamefont {C.}~\bibnamefont {Bustamante}},\ }\href@noop {} {\bibfield
  {journal} {\bibinfo  {journal} {Nature}\ }\textbf {\bibinfo {volume} {437}},\
  \bibinfo {pages} {231} (\bibinfo {year} {2005})}\BibitemShut {NoStop}%
\bibitem [{\citenamefont {Xiong}\ \emph {et~al.}(2018)\citenamefont {Xiong},
  \citenamefont {Yan}, \citenamefont {Zhou}, \citenamefont {Rehan},
  \citenamefont {Liang}, \citenamefont {Chen}, \citenamefont {Yang},
  \citenamefont {Ma}, \citenamefont {Feng},\ and\ \citenamefont
  {Vedral}}]{PhysRevLett.120.010601}%
  \BibitemOpen
  \bibfield  {author} {\bibinfo {author} {\bibfnamefont {T.~P.}\ \bibnamefont
  {Xiong}}, \bibinfo {author} {\bibfnamefont {L.~L.}\ \bibnamefont {Yan}},
  \bibinfo {author} {\bibfnamefont {F.}~\bibnamefont {Zhou}}, \bibinfo {author}
  {\bibfnamefont {K.}~\bibnamefont {Rehan}}, \bibinfo {author} {\bibfnamefont
  {D.~F.}\ \bibnamefont {Liang}}, \bibinfo {author} {\bibfnamefont
  {L.}~\bibnamefont {Chen}}, \bibinfo {author} {\bibfnamefont {W.~L.}\
  \bibnamefont {Yang}}, \bibinfo {author} {\bibfnamefont {Z.~H.}\ \bibnamefont
  {Ma}}, \bibinfo {author} {\bibfnamefont {M.}~\bibnamefont {Feng}}, \ and\
  \bibinfo {author} {\bibfnamefont {V.}~\bibnamefont {Vedral}},\ }\href
  {\doibase 10.1103/PhysRevLett.120.010601} {\bibfield  {journal} {\bibinfo
  {journal} {Phys. Rev. Lett.}\ }\textbf {\bibinfo {volume} {120}},\ \bibinfo
  {pages} {010601} (\bibinfo {year} {2018})}\BibitemShut {NoStop}%
\bibitem [{\citenamefont {Yan}\ \emph {et~al.}(2022)\citenamefont {Yan},
  \citenamefont {Zhang}, \citenamefont {Yun}, \citenamefont {Li}, \citenamefont
  {Ding}, \citenamefont {Wei}, \citenamefont {Bu}, \citenamefont {Wang},
  \citenamefont {Chen}, \citenamefont {Su}, \citenamefont {Zhou}, \citenamefont
  {Jia}, \citenamefont {Liang},\ and\ \citenamefont
  {Feng}}]{PhysRevLett.128.050603}%
  \BibitemOpen
  \bibfield  {author} {\bibinfo {author} {\bibfnamefont {L.-L.}\ \bibnamefont
  {Yan}}, \bibinfo {author} {\bibfnamefont {J.-W.}\ \bibnamefont {Zhang}},
  \bibinfo {author} {\bibfnamefont {M.-R.}\ \bibnamefont {Yun}}, \bibinfo
  {author} {\bibfnamefont {J.-C.}\ \bibnamefont {Li}}, \bibinfo {author}
  {\bibfnamefont {G.-Y.}\ \bibnamefont {Ding}}, \bibinfo {author}
  {\bibfnamefont {J.-F.}\ \bibnamefont {Wei}}, \bibinfo {author} {\bibfnamefont
  {J.-T.}\ \bibnamefont {Bu}}, \bibinfo {author} {\bibfnamefont
  {B.}~\bibnamefont {Wang}}, \bibinfo {author} {\bibfnamefont {L.}~\bibnamefont
  {Chen}}, \bibinfo {author} {\bibfnamefont {S.-L.}\ \bibnamefont {Su}},
  \bibinfo {author} {\bibfnamefont {F.}~\bibnamefont {Zhou}}, \bibinfo {author}
  {\bibfnamefont {Y.}~\bibnamefont {Jia}}, \bibinfo {author} {\bibfnamefont
  {E.-J.}\ \bibnamefont {Liang}}, \ and\ \bibinfo {author} {\bibfnamefont
  {M.}~\bibnamefont {Feng}},\ }\href {\doibase 10.1103/PhysRevLett.128.050603}
  {\bibfield  {journal} {\bibinfo  {journal} {Phys. Rev. Lett.}\ }\textbf
  {\bibinfo {volume} {128}},\ \bibinfo {pages} {050603} (\bibinfo {year}
  {2022})}\BibitemShut {NoStop}%
\bibitem [{\citenamefont {Verley}\ \emph {et~al.}(2014)\citenamefont {Verley},
  \citenamefont {Esposito}, \citenamefont {Willaert},\ and\ \citenamefont
  {Van~den Broeck}}]{verley2014unlikely}%
  \BibitemOpen
  \bibfield  {author} {\bibinfo {author} {\bibfnamefont {G.}~\bibnamefont
  {Verley}}, \bibinfo {author} {\bibfnamefont {M.}~\bibnamefont {Esposito}},
  \bibinfo {author} {\bibfnamefont {T.}~\bibnamefont {Willaert}}, \ and\
  \bibinfo {author} {\bibfnamefont {C.}~\bibnamefont {Van~den Broeck}},\
  }\href@noop {} {\bibfield  {journal} {\bibinfo  {journal} {Nature
  Communications}\ }\textbf {\bibinfo {volume} {5}},\ \bibinfo {pages} {4721}
  (\bibinfo {year} {2014})}\BibitemShut {NoStop}%
\bibitem [{\citenamefont {Cleuren}\ \emph {et~al.}(2015)\citenamefont
  {Cleuren}, \citenamefont {Rutten},\ and\ \citenamefont {Van~den
  Broeck}}]{cleuren2015universality}%
  \BibitemOpen
  \bibfield  {author} {\bibinfo {author} {\bibfnamefont {B.}~\bibnamefont
  {Cleuren}}, \bibinfo {author} {\bibfnamefont {B.}~\bibnamefont {Rutten}}, \
  and\ \bibinfo {author} {\bibfnamefont {C.}~\bibnamefont {Van~den Broeck}},\
  }\href@noop {} {\bibfield  {journal} {\bibinfo  {journal} {The European
  Physical Journal Special Topics}\ }\textbf {\bibinfo {volume} {224}},\
  \bibinfo {pages} {879} (\bibinfo {year} {2015})}\BibitemShut {NoStop}%
\bibitem [{\citenamefont {Van~den Broeck}(2005)}]{van2005thermodynamic}%
  \BibitemOpen
  \bibfield  {author} {\bibinfo {author} {\bibfnamefont {C.}~\bibnamefont
  {Van~den Broeck}},\ }\href@noop {} {\bibfield  {journal} {\bibinfo  {journal}
  {Physical Review Letters}\ }\textbf {\bibinfo {volume} {95}},\ \bibinfo
  {pages} {190602} (\bibinfo {year} {2005})}\BibitemShut {NoStop}%
\bibitem [{\citenamefont {Seifert}(2011)}]{seifert2011efficiency}%
  \BibitemOpen
  \bibfield  {author} {\bibinfo {author} {\bibfnamefont {U.}~\bibnamefont
  {Seifert}},\ }\href@noop {} {\bibfield  {journal} {\bibinfo  {journal}
  {Physical Review Letters}\ }\textbf {\bibinfo {volume} {106}},\ \bibinfo
  {pages} {020601} (\bibinfo {year} {2011})}\BibitemShut {NoStop}%
\bibitem [{\citenamefont {Golubeva}\ and\ \citenamefont
  {Imparato}(2012)}]{golubeva2012efficiency}%
  \BibitemOpen
  \bibfield  {author} {\bibinfo {author} {\bibfnamefont {N.}~\bibnamefont
  {Golubeva}}\ and\ \bibinfo {author} {\bibfnamefont {A.}~\bibnamefont
  {Imparato}},\ }\href@noop {} {\bibfield  {journal} {\bibinfo  {journal}
  {Physical Review Letters}\ }\textbf {\bibinfo {volume} {109}},\ \bibinfo
  {pages} {190602} (\bibinfo {year} {2012})}\BibitemShut {NoStop}%
\bibitem [{\citenamefont {Proesmans}\ \emph
  {et~al.}(2016{\natexlab{a}})\citenamefont {Proesmans}, \citenamefont
  {Cleuren},\ and\ \citenamefont {Van~den Broeck}}]{karel2016}%
  \BibitemOpen
  \bibfield  {author} {\bibinfo {author} {\bibfnamefont {K.}~\bibnamefont
  {Proesmans}}, \bibinfo {author} {\bibfnamefont {B.}~\bibnamefont {Cleuren}},
  \ and\ \bibinfo {author} {\bibfnamefont {C.}~\bibnamefont {Van~den Broeck}},\
  }\href {https://journals.aps.org/prl/abstract/10.1103/PhysRevLett.116.220601}
  {\bibfield  {journal} {\bibinfo  {journal} {Physical review letters}\
  }\textbf {\bibinfo {volume} {116}},\ \bibinfo {pages} {220601} (\bibinfo
  {year} {2016}{\natexlab{a}})}\BibitemShut {NoStop}%
\bibitem [{\citenamefont {Ciliberto}(2017)}]{ciliberto2017experiments}%
  \BibitemOpen
  \bibfield  {author} {\bibinfo {author} {\bibfnamefont {S.}~\bibnamefont
  {Ciliberto}},\ }\href@noop {} {\bibfield  {journal} {\bibinfo  {journal}
  {Physical Review X}\ }\textbf {\bibinfo {volume} {7}},\ \bibinfo {pages}
  {021051} (\bibinfo {year} {2017})}\BibitemShut {NoStop}%
\bibitem [{\citenamefont {Bonan{\c{c}}a}(2019)}]{bonanca2019}%
  \BibitemOpen
  \bibfield  {author} {\bibinfo {author} {\bibfnamefont {M.~V.~S.}\
  \bibnamefont {Bonan{\c{c}}a}},\ }\href {\doibase 10.1088/1742-5468/ab4e92}
  {\bibfield  {journal} {\bibinfo  {journal} {Journal of Statistical Mechanics:
  Theory and Experiment}\ }\textbf {\bibinfo {volume} {2019}},\ \bibinfo
  {pages} {123203} (\bibinfo {year} {2019})}\BibitemShut {NoStop}%
\bibitem [{\citenamefont {Campisi}\ and\ \citenamefont
  {Fazio}(2016)}]{campisi2016power}%
  \BibitemOpen
  \bibfield  {author} {\bibinfo {author} {\bibfnamefont {M.}~\bibnamefont
  {Campisi}}\ and\ \bibinfo {author} {\bibfnamefont {R.}~\bibnamefont
  {Fazio}},\ }\href@noop {} {\bibfield  {journal} {\bibinfo  {journal} {Nature
  communications}\ }\textbf {\bibinfo {volume} {7}},\ \bibinfo {pages} {1}
  (\bibinfo {year} {2016})}\BibitemShut {NoStop}%
\bibitem [{\citenamefont {Proesmans}\ \emph
  {et~al.}(2016{\natexlab{b}})\citenamefont {Proesmans}, \citenamefont
  {Cleuren},\ and\ \citenamefont {Van~den Broeck}}]{karel2016prl}%
  \BibitemOpen
  \bibfield  {author} {\bibinfo {author} {\bibfnamefont {K.}~\bibnamefont
  {Proesmans}}, \bibinfo {author} {\bibfnamefont {B.}~\bibnamefont {Cleuren}},
  \ and\ \bibinfo {author} {\bibfnamefont {C.}~\bibnamefont {Van~den Broeck}},\
  }\href {https://journals.aps.org/prl/abstract/10.1103/PhysRevLett.116.220601}
  {\bibfield  {journal} {\bibinfo  {journal} {Physical review letters}\
  }\textbf {\bibinfo {volume} {116}},\ \bibinfo {pages} {220601} (\bibinfo
  {year} {2016}{\natexlab{b}})}\BibitemShut {NoStop}%
\bibitem [{\citenamefont {Mamede}\ \emph {et~al.}(2022)\citenamefont {Mamede},
  \citenamefont {Harunari}, \citenamefont {Akasaki}, \citenamefont
  {Proesmans},\ and\ \citenamefont {Fiore}}]{mamede2021obtaining}%
  \BibitemOpen
  \bibfield  {author} {\bibinfo {author} {\bibfnamefont {I.~N.}\ \bibnamefont
  {Mamede}}, \bibinfo {author} {\bibfnamefont {P.~E.}\ \bibnamefont
  {Harunari}}, \bibinfo {author} {\bibfnamefont {B.~A.~N.}\ \bibnamefont
  {Akasaki}}, \bibinfo {author} {\bibfnamefont {K.}~\bibnamefont {Proesmans}},
  \ and\ \bibinfo {author} {\bibfnamefont {C.~E.}\ \bibnamefont {Fiore}},\
  }\href {\doibase 10.1103/PhysRevE.105.024106} {\bibfield  {journal} {\bibinfo
   {journal} {Phys. Rev. E}\ }\textbf {\bibinfo {volume} {105}},\ \bibinfo
  {pages} {024106} (\bibinfo {year} {2022})}\BibitemShut {NoStop}%
\bibitem [{\citenamefont {Noa}\ \emph {et~al.}(2021)\citenamefont {Noa},
  \citenamefont {Stable}, \citenamefont {Oropesa}, \citenamefont {Rosas},\ and\
  \citenamefont {Fiore}}]{noa2021efficient}%
  \BibitemOpen
  \bibfield  {author} {\bibinfo {author} {\bibfnamefont {C.~E.~F.}\
  \bibnamefont {Noa}}, \bibinfo {author} {\bibfnamefont {A.~L.~L.}\
  \bibnamefont {Stable}}, \bibinfo {author} {\bibfnamefont {W.~G.~C.}\
  \bibnamefont {Oropesa}}, \bibinfo {author} {\bibfnamefont {A.}~\bibnamefont
  {Rosas}}, \ and\ \bibinfo {author} {\bibfnamefont {C.~E.}\ \bibnamefont
  {Fiore}},\ }\href {\doibase 10.1103/PhysRevResearch.3.043152} {\bibfield
  {journal} {\bibinfo  {journal} {Phys. Rev. Research}\ }\textbf {\bibinfo
  {volume} {3}},\ \bibinfo {pages} {043152} (\bibinfo {year}
  {2021})}\BibitemShut {NoStop}%
\bibitem [{\citenamefont {Harunari}\ \emph {et~al.}(2021)\citenamefont
  {Harunari}, \citenamefont {Filho}, \citenamefont {Fiore},\ and\ \citenamefont
  {Rosas}}]{harunari2020maximal}%
  \BibitemOpen
  \bibfield  {author} {\bibinfo {author} {\bibfnamefont {P.~E.}\ \bibnamefont
  {Harunari}}, \bibinfo {author} {\bibfnamefont {F.~S.}\ \bibnamefont {Filho}},
  \bibinfo {author} {\bibfnamefont {C.~E.}\ \bibnamefont {Fiore}}, \ and\
  \bibinfo {author} {\bibfnamefont {A.}~\bibnamefont {Rosas}},\ }\href
  {\doibase 10.1103/PhysRevResearch.3.023194} {\bibfield  {journal} {\bibinfo
  {journal} {Phys. Rev. Research}\ }\textbf {\bibinfo {volume} {3}},\ \bibinfo
  {pages} {023194} (\bibinfo {year} {2021})}\BibitemShut {NoStop}%
\bibitem [{\citenamefont {Gu\'ery-Odelin}\ \emph {et~al.}(2019)\citenamefont
  {Gu\'ery-Odelin}, \citenamefont {Ruschhaupt}, \citenamefont {Kiely},
  \citenamefont {Torrontegui}, \citenamefont {Mart\'{\i}nez-Garaot},\ and\
  \citenamefont {Muga}}]{RevModPhys.91.045001}%
  \BibitemOpen
  \bibfield  {author} {\bibinfo {author} {\bibfnamefont {D.}~\bibnamefont
  {Gu\'ery-Odelin}}, \bibinfo {author} {\bibfnamefont {A.}~\bibnamefont
  {Ruschhaupt}}, \bibinfo {author} {\bibfnamefont {A.}~\bibnamefont {Kiely}},
  \bibinfo {author} {\bibfnamefont {E.}~\bibnamefont {Torrontegui}}, \bibinfo
  {author} {\bibfnamefont {S.}~\bibnamefont {Mart\'{\i}nez-Garaot}}, \ and\
  \bibinfo {author} {\bibfnamefont {J.~G.}\ \bibnamefont {Muga}},\ }\href
  {\doibase 10.1103/RevModPhys.91.045001} {\bibfield  {journal} {\bibinfo
  {journal} {Rev. Mod. Phys.}\ }\textbf {\bibinfo {volume} {91}},\ \bibinfo
  {pages} {045001} (\bibinfo {year} {2019})}\BibitemShut {NoStop}%
\bibitem [{\citenamefont {Deffner}\ and\ \citenamefont
  {Bonan{\c{c}}a}(2020)}]{deffner2020thermodynamic}%
  \BibitemOpen
  \bibfield  {author} {\bibinfo {author} {\bibfnamefont {S.}~\bibnamefont
  {Deffner}}\ and\ \bibinfo {author} {\bibfnamefont {M.~V.}\ \bibnamefont
  {Bonan{\c{c}}a}},\ }\href@noop {} {\bibfield  {journal} {\bibinfo  {journal}
  {EPL (Europhysics Letters)}\ }\textbf {\bibinfo {volume} {131}},\ \bibinfo
  {pages} {20001} (\bibinfo {year} {2020})}\BibitemShut {NoStop}%
\bibitem [{\citenamefont {Pancotti}\ \emph {et~al.}(2020)\citenamefont
  {Pancotti}, \citenamefont {Scandi}, \citenamefont {Mitchison},\ and\
  \citenamefont {Perarnau-Llobet}}]{pancotti2020speed}%
  \BibitemOpen
  \bibfield  {author} {\bibinfo {author} {\bibfnamefont {N.}~\bibnamefont
  {Pancotti}}, \bibinfo {author} {\bibfnamefont {M.}~\bibnamefont {Scandi}},
  \bibinfo {author} {\bibfnamefont {M.~T.}\ \bibnamefont {Mitchison}}, \ and\
  \bibinfo {author} {\bibfnamefont {M.}~\bibnamefont {Perarnau-Llobet}},\
  }\href@noop {} {\bibfield  {journal} {\bibinfo  {journal} {Physical Review
  X}\ }\textbf {\bibinfo {volume} {10}},\ \bibinfo {pages} {031015} (\bibinfo
  {year} {2020})}\BibitemShut {NoStop}%
\bibitem [{\citenamefont {Zhao}\ \emph {et~al.}(2022)\citenamefont {Zhao},
  \citenamefont {Gong},\ and\ \citenamefont {Tu}}]{2206.02337}%
  \BibitemOpen
  \bibfield  {author} {\bibinfo {author} {\bibfnamefont {X.-H.}\ \bibnamefont
  {Zhao}}, \bibinfo {author} {\bibfnamefont {Z.-N.}\ \bibnamefont {Gong}}, \
  and\ \bibinfo {author} {\bibfnamefont {Z.~C.}\ \bibnamefont {Tu}},\ }\href
  {\doibase 10.48550/ARXIV.2206.02337} {\enquote {\bibinfo {title} {Microscopic
  low-dissipation heat engine via shortcuts to adiabaticity and shortcuts to
  isothermality},}\ } (\bibinfo {year} {2022})\BibitemShut {NoStop}%
\bibitem [{\citenamefont {Gnesotto}\ \emph {et~al.}(2018)\citenamefont
  {Gnesotto}, \citenamefont {Mura}, \citenamefont {Gladrow},\ and\
  \citenamefont {Broedersz}}]{gnesotto2018broken}%
  \BibitemOpen
  \bibfield  {author} {\bibinfo {author} {\bibfnamefont {F.~S.}\ \bibnamefont
  {Gnesotto}}, \bibinfo {author} {\bibfnamefont {F.}~\bibnamefont {Mura}},
  \bibinfo {author} {\bibfnamefont {J.}~\bibnamefont {Gladrow}}, \ and\
  \bibinfo {author} {\bibfnamefont {C.~P.}\ \bibnamefont {Broedersz}},\
  }\href@noop {} {\bibfield  {journal} {\bibinfo  {journal} {Reports on
  Progress in Physics}\ }\textbf {\bibinfo {volume} {81}},\ \bibinfo {pages}
  {066601} (\bibinfo {year} {2018})}\BibitemShut {NoStop}%
\bibitem [{\citenamefont {Smith}\ and\ \citenamefont
  {Schuster}(2019)}]{smith2019public}%
  \BibitemOpen
  \bibfield  {author} {\bibinfo {author} {\bibfnamefont {P.}~\bibnamefont
  {Smith}}\ and\ \bibinfo {author} {\bibfnamefont {M.}~\bibnamefont
  {Schuster}},\ }\href@noop {} {\bibfield  {journal} {\bibinfo  {journal}
  {Current Biology}\ }\textbf {\bibinfo {volume} {29}},\ \bibinfo {pages}
  {R442} (\bibinfo {year} {2019})}\BibitemShut {NoStop}%
\bibitem [{\citenamefont {Lynn}\ \emph {et~al.}(2021)\citenamefont {Lynn},
  \citenamefont {Cornblath}, \citenamefont {Papadopoulos}, \citenamefont
  {Bertolero},\ and\ \citenamefont {Bassett}}]{lynn2021broken}%
  \BibitemOpen
  \bibfield  {author} {\bibinfo {author} {\bibfnamefont {C.~W.}\ \bibnamefont
  {Lynn}}, \bibinfo {author} {\bibfnamefont {E.~J.}\ \bibnamefont {Cornblath}},
  \bibinfo {author} {\bibfnamefont {L.}~\bibnamefont {Papadopoulos}}, \bibinfo
  {author} {\bibfnamefont {M.~A.}\ \bibnamefont {Bertolero}}, \ and\ \bibinfo
  {author} {\bibfnamefont {D.~S.}\ \bibnamefont {Bassett}},\ }\href@noop {}
  {\bibfield  {journal} {\bibinfo  {journal} {Proceedings of the National
  Academy of Sciences}\ }\textbf {\bibinfo {volume} {118}},\ \bibinfo {pages}
  {e2109889118} (\bibinfo {year} {2021})}\BibitemShut {NoStop}%
\bibitem [{\citenamefont {Rao}\ and\ \citenamefont
  {Esposito}(2016)}]{rao2016nonequilibrium}%
  \BibitemOpen
  \bibfield  {author} {\bibinfo {author} {\bibfnamefont {R.}~\bibnamefont
  {Rao}}\ and\ \bibinfo {author} {\bibfnamefont {M.}~\bibnamefont {Esposito}},\
  }\href@noop {} {\bibfield  {journal} {\bibinfo  {journal} {Physical Review
  X}\ }\textbf {\bibinfo {volume} {6}},\ \bibinfo {pages} {041064} (\bibinfo
  {year} {2016})}\BibitemShut {NoStop}%
\bibitem [{\citenamefont {Busiello}\ \emph {et~al.}(2021)\citenamefont
  {Busiello}, \citenamefont {Liang}, \citenamefont {Piazza},\ and\
  \citenamefont {De~Los~Rios}}]{busiello2021dissipation}%
  \BibitemOpen
  \bibfield  {author} {\bibinfo {author} {\bibfnamefont {D.~M.}\ \bibnamefont
  {Busiello}}, \bibinfo {author} {\bibfnamefont {S.}~\bibnamefont {Liang}},
  \bibinfo {author} {\bibfnamefont {F.}~\bibnamefont {Piazza}}, \ and\ \bibinfo
  {author} {\bibfnamefont {P.}~\bibnamefont {De~Los~Rios}},\ }\href@noop {}
  {\bibfield  {journal} {\bibinfo  {journal} {Communications Chemistry}\
  }\textbf {\bibinfo {volume} {4}},\ \bibinfo {pages} {1} (\bibinfo {year}
  {2021})}\BibitemShut {NoStop}%
\bibitem [{\citenamefont {Dass}\ \emph {et~al.}(2021)\citenamefont {Dass},
  \citenamefont {Georgelin}, \citenamefont {Westall}, \citenamefont {Foucher},
  \citenamefont {De~Los~Rios}, \citenamefont {Busiello}, \citenamefont
  {Liang},\ and\ \citenamefont {Piazza}}]{dass2021equilibrium}%
  \BibitemOpen
  \bibfield  {author} {\bibinfo {author} {\bibfnamefont {A.~V.}\ \bibnamefont
  {Dass}}, \bibinfo {author} {\bibfnamefont {T.}~\bibnamefont {Georgelin}},
  \bibinfo {author} {\bibfnamefont {F.}~\bibnamefont {Westall}}, \bibinfo
  {author} {\bibfnamefont {F.}~\bibnamefont {Foucher}}, \bibinfo {author}
  {\bibfnamefont {P.}~\bibnamefont {De~Los~Rios}}, \bibinfo {author}
  {\bibfnamefont {D.~M.}\ \bibnamefont {Busiello}}, \bibinfo {author}
  {\bibfnamefont {S.}~\bibnamefont {Liang}}, \ and\ \bibinfo {author}
  {\bibfnamefont {F.}~\bibnamefont {Piazza}},\ }\href@noop {} {\bibfield
  {journal} {\bibinfo  {journal} {Nature communications}\ }\textbf {\bibinfo
  {volume} {12}},\ \bibinfo {pages} {1} (\bibinfo {year} {2021})}\BibitemShut
  {NoStop}%
\bibitem [{\citenamefont {Bonifazi}\ \emph {et~al.}(2009)\citenamefont
  {Bonifazi}, \citenamefont {Goldin}, \citenamefont {Picardo}, \citenamefont
  {Jorquera}, \citenamefont {Cattani}, \citenamefont {Bianconi}, \citenamefont
  {Represa}, \citenamefont {Ben-Ari},\ and\ \citenamefont
  {Cossart}}]{bonifazi2009gabaergic}%
  \BibitemOpen
  \bibfield  {author} {\bibinfo {author} {\bibfnamefont {P.}~\bibnamefont
  {Bonifazi}}, \bibinfo {author} {\bibfnamefont {M.}~\bibnamefont {Goldin}},
  \bibinfo {author} {\bibfnamefont {M.~A.}\ \bibnamefont {Picardo}}, \bibinfo
  {author} {\bibfnamefont {I.}~\bibnamefont {Jorquera}}, \bibinfo {author}
  {\bibfnamefont {A.}~\bibnamefont {Cattani}}, \bibinfo {author} {\bibfnamefont
  {G.}~\bibnamefont {Bianconi}}, \bibinfo {author} {\bibfnamefont
  {A.}~\bibnamefont {Represa}}, \bibinfo {author} {\bibfnamefont
  {Y.}~\bibnamefont {Ben-Ari}}, \ and\ \bibinfo {author} {\bibfnamefont
  {R.}~\bibnamefont {Cossart}},\ }\href@noop {} {\bibfield  {journal} {\bibinfo
   {journal} {Science}\ }\textbf {\bibinfo {volume} {326}},\ \bibinfo {pages}
  {1419} (\bibinfo {year} {2009})}\BibitemShut {NoStop}%
\bibitem [{\citenamefont {Schneidman}\ \emph {et~al.}(2006)\citenamefont
  {Schneidman}, \citenamefont {Berry}, \citenamefont {Segev},\ and\
  \citenamefont {Bialek}}]{schneidman2006weak}%
  \BibitemOpen
  \bibfield  {author} {\bibinfo {author} {\bibfnamefont {E.}~\bibnamefont
  {Schneidman}}, \bibinfo {author} {\bibfnamefont {M.~J.}\ \bibnamefont
  {Berry}}, \bibinfo {author} {\bibfnamefont {R.}~\bibnamefont {Segev}}, \ and\
  \bibinfo {author} {\bibfnamefont {W.}~\bibnamefont {Bialek}},\ }\href@noop {}
  {\bibfield  {journal} {\bibinfo  {journal} {Nature}\ }\textbf {\bibinfo
  {volume} {440}},\ \bibinfo {pages} {1007} (\bibinfo {year}
  {2006})}\BibitemShut {NoStop}%
\bibitem [{\citenamefont {Buzs{\'a}ki}\ and\ \citenamefont
  {Mizuseki}(2014)}]{buzsaki2014log}%
  \BibitemOpen
  \bibfield  {author} {\bibinfo {author} {\bibfnamefont {G.}~\bibnamefont
  {Buzs{\'a}ki}}\ and\ \bibinfo {author} {\bibfnamefont {K.}~\bibnamefont
  {Mizuseki}},\ }\href@noop {} {\bibfield  {journal} {\bibinfo  {journal}
  {Nature Reviews Neuroscience}\ }\textbf {\bibinfo {volume} {15}},\ \bibinfo
  {pages} {264} (\bibinfo {year} {2014})}\BibitemShut {NoStop}%
\bibitem [{\citenamefont {Gal}\ \emph {et~al.}(2017)\citenamefont {Gal},
  \citenamefont {London}, \citenamefont {Globerson}, \citenamefont {Ramaswamy},
  \citenamefont {Reimann}, \citenamefont {Muller}, \citenamefont {Markram},\
  and\ \citenamefont {Segev}}]{gal2017rich}%
  \BibitemOpen
  \bibfield  {author} {\bibinfo {author} {\bibfnamefont {E.}~\bibnamefont
  {Gal}}, \bibinfo {author} {\bibfnamefont {M.}~\bibnamefont {London}},
  \bibinfo {author} {\bibfnamefont {A.}~\bibnamefont {Globerson}}, \bibinfo
  {author} {\bibfnamefont {S.}~\bibnamefont {Ramaswamy}}, \bibinfo {author}
  {\bibfnamefont {M.~W.}\ \bibnamefont {Reimann}}, \bibinfo {author}
  {\bibfnamefont {E.}~\bibnamefont {Muller}}, \bibinfo {author} {\bibfnamefont
  {H.}~\bibnamefont {Markram}}, \ and\ \bibinfo {author} {\bibfnamefont
  {I.}~\bibnamefont {Segev}},\ }\href@noop {} {\bibfield  {journal} {\bibinfo
  {journal} {Nature neuroscience}\ }\textbf {\bibinfo {volume} {20}},\ \bibinfo
  {pages} {1004} (\bibinfo {year} {2017})}\BibitemShut {NoStop}%
\bibitem [{\citenamefont {T{\"o}njes}\ \emph {et~al.}(2021)\citenamefont
  {T{\"o}njes}, \citenamefont {Fiore},\ and\ \citenamefont
  {Pereira}}]{tonjes2021coherence}%
  \BibitemOpen
  \bibfield  {author} {\bibinfo {author} {\bibfnamefont {R.}~\bibnamefont
  {T{\"o}njes}}, \bibinfo {author} {\bibfnamefont {C.~E.}\ \bibnamefont
  {Fiore}}, \ and\ \bibinfo {author} {\bibfnamefont {T.}~\bibnamefont
  {Pereira}},\ }\href@noop {} {\bibfield  {journal} {\bibinfo  {journal}
  {Nature Communications}\ }\textbf {\bibinfo {volume} {12}},\ \bibinfo {pages}
  {1} (\bibinfo {year} {2021})}\BibitemShut {NoStop}%
\bibitem [{\citenamefont {Mukherjee}\ and\ \citenamefont
  {Divakaran}(2021)}]{mukherjee2021manybody}%
  \BibitemOpen
  \bibfield  {author} {\bibinfo {author} {\bibfnamefont {V.}~\bibnamefont
  {Mukherjee}}\ and\ \bibinfo {author} {\bibfnamefont {U.}~\bibnamefont
  {Divakaran}},\ }\href@noop {} {\bibfield  {journal} {\bibinfo  {journal}
  {Journal of Physics: Condensed Matter}\ }\textbf {\bibinfo {volume} {33}}
  (\bibinfo {year} {2021})}\BibitemShut {NoStop}%
\bibitem [{\citenamefont {Niedenzu}\ and\ \citenamefont
  {Kurizki}(2018)}]{niedenzu2018cooperative}%
  \BibitemOpen
  \bibfield  {author} {\bibinfo {author} {\bibfnamefont {W.}~\bibnamefont
  {Niedenzu}}\ and\ \bibinfo {author} {\bibfnamefont {G.}~\bibnamefont
  {Kurizki}},\ }\href {\doibase 10.1088/1367-2630/aaed55} {\bibfield  {journal}
  {\bibinfo  {journal} {New Journal of Physics}\ }\textbf {\bibinfo {volume}
  {20}},\ \bibinfo {pages} {113038} (\bibinfo {year} {2018})}\BibitemShut
  {NoStop}%
\bibitem [{\citenamefont {Kolisnyk}\ and\ \citenamefont
  {Schaller}(2023)}]{PhysRevApplied.19.034023}%
  \BibitemOpen
  \bibfield  {author} {\bibinfo {author} {\bibfnamefont {D.}~\bibnamefont
  {Kolisnyk}}\ and\ \bibinfo {author} {\bibfnamefont {G.}~\bibnamefont
  {Schaller}},\ }\href {\doibase 10.1103/PhysRevApplied.19.034023} {\bibfield
  {journal} {\bibinfo  {journal} {Phys. Rev. Appl.}\ }\textbf {\bibinfo
  {volume} {19}},\ \bibinfo {pages} {034023} (\bibinfo {year}
  {2023})}\BibitemShut {NoStop}%
\bibitem [{\citenamefont {C.}\ \emph {et~al.}(2020)\citenamefont {C.},
  \citenamefont {I.},\ and\ \citenamefont {F.}}]{latune2020collective}%
  \BibitemOpen
  \bibfield  {author} {\bibinfo {author} {\bibfnamefont {L.}~\bibnamefont {C.},
  \bibfnamefont {L.}}, \bibinfo {author} {\bibfnamefont {S.}~\bibnamefont
  {I.}}, \ and\ \bibinfo {author} {\bibfnamefont {P.}~\bibnamefont {F.}},\
  }\href {\doibase 10.1088/1367-2630/aba463} {\bibfield  {journal} {\bibinfo
  {journal} {New Journal of Physics}\ }\textbf {\bibinfo {volume} {22}},\
  \bibinfo {pages} {083049} (\bibinfo {year} {2020})}\BibitemShut {NoStop}%
\bibitem [{\citenamefont {Kamimura}\ \emph {et~al.}(2022)\citenamefont
  {Kamimura}, \citenamefont {Hakoshimam}, \citenamefont {Matsuzaki},
  \citenamefont {Yoshida},\ and\ \citenamefont
  {Tokura}}]{kamimura2022collective}%
  \BibitemOpen
  \bibfield  {author} {\bibinfo {author} {\bibfnamefont {S.}~\bibnamefont
  {Kamimura}}, \bibinfo {author} {\bibfnamefont {H.}~\bibnamefont
  {Hakoshimam}}, \bibinfo {author} {\bibfnamefont {Y.}~\bibnamefont
  {Matsuzaki}}, \bibinfo {author} {\bibfnamefont {K.}~\bibnamefont {Yoshida}},
  \ and\ \bibinfo {author} {\bibfnamefont {Y.}~\bibnamefont {Tokura}},\ }\href
  {\doibase 10.1103/PhysRevLett.128.180602} {\bibfield  {journal} {\bibinfo
  {journal} {Physical Review Letter}\ }\textbf {\bibinfo {volume} {128}},\
  \bibinfo {pages} {180602} (\bibinfo {year} {2022})}\BibitemShut {NoStop}%
\bibitem [{\citenamefont {Macovei}(2022)}]{mavocei2022performance}%
  \BibitemOpen
  \bibfield  {author} {\bibinfo {author} {\bibfnamefont {M.}~\bibnamefont
  {Macovei}, \bibfnamefont {A.}},\ }\href {\doibase
  10.1103/PhysRevA.105.043708} {\bibfield  {journal} {\bibinfo  {journal}
  {Physical Review A}\ }\textbf {\bibinfo {volume} {105}},\ \bibinfo {pages}
  {043708} (\bibinfo {year} {2022})}\BibitemShut {NoStop}%
\bibitem [{\citenamefont {Vroylandt}\ \emph {et~al.}(2017)\citenamefont
  {Vroylandt}, \citenamefont {Esposito},\ and\ \citenamefont
  {Verley}}]{gatien}%
  \BibitemOpen
  \bibfield  {author} {\bibinfo {author} {\bibfnamefont {H.}~\bibnamefont
  {Vroylandt}}, \bibinfo {author} {\bibfnamefont {M.}~\bibnamefont {Esposito}},
  \ and\ \bibinfo {author} {\bibfnamefont {G.}~\bibnamefont {Verley}},\ }\href
  {\doibase 10.1209/0295-5075/120/30009} {\bibfield  {journal} {\bibinfo
  {journal} {{EPL} (Europhysics Letters)}\ }\textbf {\bibinfo {volume} {120}},\
  \bibinfo {pages} {30009} (\bibinfo {year} {2017})}\BibitemShut {NoStop}%
\bibitem [{\citenamefont {Herpich}\ \emph {et~al.}(2018)\citenamefont
  {Herpich}, \citenamefont {Thingna},\ and\ \citenamefont
  {Esposito}}]{herpich}%
  \BibitemOpen
  \bibfield  {author} {\bibinfo {author} {\bibfnamefont {T.}~\bibnamefont
  {Herpich}}, \bibinfo {author} {\bibfnamefont {J.}~\bibnamefont {Thingna}}, \
  and\ \bibinfo {author} {\bibfnamefont {M.}~\bibnamefont {Esposito}},\ }\href
  {\doibase 10.1103/PhysRevX.8.031056} {\bibfield  {journal} {\bibinfo
  {journal} {Phys. Rev. X}\ }\textbf {\bibinfo {volume} {8}},\ \bibinfo {pages}
  {031056} (\bibinfo {year} {2018})}\BibitemShut {NoStop}%
\bibitem [{\citenamefont {Herpich}\ and\ \citenamefont
  {Esposito}(2019)}]{herpich2}%
  \BibitemOpen
  \bibfield  {author} {\bibinfo {author} {\bibfnamefont {T.}~\bibnamefont
  {Herpich}}\ and\ \bibinfo {author} {\bibfnamefont {M.}~\bibnamefont
  {Esposito}},\ }\href {\doibase 10.1103/PhysRevE.99.022135} {\bibfield
  {journal} {\bibinfo  {journal} {Phys. Rev. E}\ }\textbf {\bibinfo {volume}
  {99}},\ \bibinfo {pages} {022135} (\bibinfo {year} {2019})}\BibitemShut
  {NoStop}%
\bibitem [{\citenamefont {Su{\~n}{\'e}}\ and\ \citenamefont
  {Imparato}(2019)}]{sune2019out}%
  \BibitemOpen
  \bibfield  {author} {\bibinfo {author} {\bibfnamefont {M.}~\bibnamefont
  {Su{\~n}{\'e}}}\ and\ \bibinfo {author} {\bibfnamefont {A.}~\bibnamefont
  {Imparato}},\ }\href@noop {} {\bibfield  {journal} {\bibinfo  {journal}
  {Physical Review Letters}\ }\textbf {\bibinfo {volume} {123}},\ \bibinfo
  {pages} {070601} (\bibinfo {year} {2019})}\BibitemShut {NoStop}%
\bibitem [{\citenamefont {Yeomans}(1992)}]{yeomans1992statistical}%
  \BibitemOpen
  \bibfield  {author} {\bibinfo {author} {\bibfnamefont {J.~M.}\ \bibnamefont
  {Yeomans}},\ }\href@noop {} {\emph {\bibinfo {title} {Statistical mechanics
  of phase transitions}}}\ (\bibinfo  {publisher} {Clarendon Press},\ \bibinfo
  {year} {1992})\BibitemShut {NoStop}%
\bibitem [{\citenamefont {Wu}(1982)}]{RevModPhys.54.235}%
  \BibitemOpen
  \bibfield  {author} {\bibinfo {author} {\bibfnamefont {F.~Y.}\ \bibnamefont
  {Wu}},\ }\href {\doibase 10.1103/RevModPhys.54.235} {\bibfield  {journal}
  {\bibinfo  {journal} {Rev. Mod. Phys.}\ }\textbf {\bibinfo {volume} {54}},\
  \bibinfo {pages} {235} (\bibinfo {year} {1982})}\BibitemShut {NoStop}%
\bibitem [{\citenamefont {Blume}\ \emph {et~al.}(1971)\citenamefont {Blume},
  \citenamefont {Emery},\ and\ \citenamefont {Griffiths}}]{PhysRevA.4.1071}%
  \BibitemOpen
  \bibfield  {author} {\bibinfo {author} {\bibfnamefont {M.}~\bibnamefont
  {Blume}}, \bibinfo {author} {\bibfnamefont {V.~J.}\ \bibnamefont {Emery}}, \
  and\ \bibinfo {author} {\bibfnamefont {R.~B.}\ \bibnamefont {Griffiths}},\
  }\href {\doibase 10.1103/PhysRevA.4.1071} {\bibfield  {journal} {\bibinfo
  {journal} {Phys. Rev. A}\ }\textbf {\bibinfo {volume} {4}},\ \bibinfo {pages}
  {1071} (\bibinfo {year} {1971})}\BibitemShut {NoStop}%
\bibitem [{\citenamefont {Hoston}\ and\ \citenamefont
  {Berker}(1991)}]{PhysRevLett.67.1027}%
  \BibitemOpen
  \bibfield  {author} {\bibinfo {author} {\bibfnamefont {W.}~\bibnamefont
  {Hoston}}\ and\ \bibinfo {author} {\bibfnamefont {A.~N.}\ \bibnamefont
  {Berker}},\ }\href {\doibase 10.1103/PhysRevLett.67.1027} {\bibfield
  {journal} {\bibinfo  {journal} {Phys. Rev. Lett.}\ }\textbf {\bibinfo
  {volume} {67}},\ \bibinfo {pages} {1027} (\bibinfo {year}
  {1991})}\BibitemShut {NoStop}%
\bibitem [{\citenamefont {Busiello}\ \emph {et~al.}(2020)\citenamefont
  {Busiello}, \citenamefont {Gupta},\ and\ \citenamefont
  {Maritan}}]{busiello2020coarse}%
  \BibitemOpen
  \bibfield  {author} {\bibinfo {author} {\bibfnamefont {D.~M.}\ \bibnamefont
  {Busiello}}, \bibinfo {author} {\bibfnamefont {D.}~\bibnamefont {Gupta}}, \
  and\ \bibinfo {author} {\bibfnamefont {A.}~\bibnamefont {Maritan}},\
  }\href@noop {} {\bibfield  {journal} {\bibinfo  {journal} {Physical Review
  Research}\ }\textbf {\bibinfo {volume} {2}},\ \bibinfo {pages} {043257}
  (\bibinfo {year} {2020})}\BibitemShut {NoStop}%
\bibitem [{\citenamefont {Salinas}(2001)}]{salinas2001introduction}%
  \BibitemOpen
  \bibfield  {author} {\bibinfo {author} {\bibfnamefont {S.~R.}\ \bibnamefont
  {Salinas}},\ }in\ \href@noop {} {\emph {\bibinfo {booktitle} {Introduction to
  Statistical Physics}}}\ (\bibinfo  {publisher} {Springer},\ \bibinfo {year}
  {2001})\ pp.\ \bibinfo {pages} {1--17}\BibitemShut {NoStop}%
\bibitem [{\citenamefont {Liepelt}\ and\ \citenamefont
  {Lipowsky}(2007)}]{liepelt1}%
  \BibitemOpen
  \bibfield  {author} {\bibinfo {author} {\bibfnamefont {S.}~\bibnamefont
  {Liepelt}}\ and\ \bibinfo {author} {\bibfnamefont {R.}~\bibnamefont
  {Lipowsky}},\ }\href@noop {} {\bibfield  {journal} {\bibinfo  {journal}
  {Phys. Rev. Lett.}\ }\textbf {\bibinfo {volume} {98}},\ \bibinfo {pages}
  {258102} (\bibinfo {year} {2007})}\BibitemShut {NoStop}%
\bibitem [{\citenamefont {Liepelt}\ and\ \citenamefont
  {Lipowsky}(2009)}]{liepelt2}%
  \BibitemOpen
  \bibfield  {author} {\bibinfo {author} {\bibfnamefont {S.}~\bibnamefont
  {Liepelt}}\ and\ \bibinfo {author} {\bibfnamefont {R.}~\bibnamefont
  {Lipowsky}},\ }\href {\doibase 10.1103/PhysRevE.79.011917} {\bibfield
  {journal} {\bibinfo  {journal} {Phys. Rev. E}\ }\textbf {\bibinfo {volume}
  {79}},\ \bibinfo {pages} {011917} (\bibinfo {year} {2009})}\BibitemShut
  {NoStop}%
\bibitem [{\citenamefont {Berton}\ \emph {et~al.}(2020)\citenamefont {Berton},
  \citenamefont {Busiello}, \citenamefont {Zamuner}, \citenamefont {Solari},
  \citenamefont {Scopelliti}, \citenamefont {Fadaei-Tirani}, \citenamefont
  {Severin},\ and\ \citenamefont {Pezzato}}]{berton2020thermodynamics}%
  \BibitemOpen
  \bibfield  {author} {\bibinfo {author} {\bibfnamefont {C.}~\bibnamefont
  {Berton}}, \bibinfo {author} {\bibfnamefont {D.~M.}\ \bibnamefont
  {Busiello}}, \bibinfo {author} {\bibfnamefont {S.}~\bibnamefont {Zamuner}},
  \bibinfo {author} {\bibfnamefont {E.}~\bibnamefont {Solari}}, \bibinfo
  {author} {\bibfnamefont {R.}~\bibnamefont {Scopelliti}}, \bibinfo {author}
  {\bibfnamefont {F.}~\bibnamefont {Fadaei-Tirani}}, \bibinfo {author}
  {\bibfnamefont {K.}~\bibnamefont {Severin}}, \ and\ \bibinfo {author}
  {\bibfnamefont {C.}~\bibnamefont {Pezzato}},\ }\href@noop {} {\bibfield
  {journal} {\bibinfo  {journal} {Chemical Science}\ }\textbf {\bibinfo
  {volume} {11}},\ \bibinfo {pages} {8457} (\bibinfo {year}
  {2020})}\BibitemShut {NoStop}%
\bibitem [{\citenamefont {De~Los~Rios}\ and\ \citenamefont
  {Barducci}(2014)}]{de2014hsp70}%
  \BibitemOpen
  \bibfield  {author} {\bibinfo {author} {\bibfnamefont {P.}~\bibnamefont
  {De~Los~Rios}}\ and\ \bibinfo {author} {\bibfnamefont {A.}~\bibnamefont
  {Barducci}},\ }\href@noop {} {\bibfield  {journal} {\bibinfo  {journal}
  {Elife}\ }\textbf {\bibinfo {volume} {3}},\ \bibinfo {pages} {e02218}
  (\bibinfo {year} {2014})}\BibitemShut {NoStop}%
\bibitem [{\citenamefont {Schnakenberg}(1976)}]{schnakenberg}%
  \BibitemOpen
  \bibfield  {author} {\bibinfo {author} {\bibfnamefont {J.}~\bibnamefont
  {Schnakenberg}},\ }\href@noop {} {\bibfield  {journal} {\bibinfo  {journal}
  {Reviews of Modern physics}\ }\textbf {\bibinfo {volume} {48}},\ \bibinfo
  {pages} {571} (\bibinfo {year} {1976})}\BibitemShut {NoStop}%
\bibitem [{\citenamefont {J{\"u}licher}\ and\ \citenamefont
  {Prost}(1995)}]{julicher1995cooperative}%
  \BibitemOpen
  \bibfield  {author} {\bibinfo {author} {\bibfnamefont {F.}~\bibnamefont
  {J{\"u}licher}}\ and\ \bibinfo {author} {\bibfnamefont {J.}~\bibnamefont
  {Prost}},\ }\href@noop {} {\bibfield  {journal} {\bibinfo  {journal}
  {Physical review letters}\ }\textbf {\bibinfo {volume} {75}},\ \bibinfo
  {pages} {2618} (\bibinfo {year} {1995})}\BibitemShut {NoStop}%
\bibitem [{\citenamefont {Tu}(2008)}]{tu2008nonequilibrium}%
  \BibitemOpen
  \bibfield  {author} {\bibinfo {author} {\bibfnamefont {Y.}~\bibnamefont
  {Tu}},\ }\href@noop {} {\bibfield  {journal} {\bibinfo  {journal}
  {Proceedings of the National Academy of Sciences}\ }\textbf {\bibinfo
  {volume} {105}},\ \bibinfo {pages} {11737} (\bibinfo {year}
  {2008})}\BibitemShut {NoStop}%
\bibitem [{\citenamefont {Horowitz}\ and\ \citenamefont
  {Esposito}(2014)}]{horowitz2014thermodynamics}%
  \BibitemOpen
  \bibfield  {author} {\bibinfo {author} {\bibfnamefont {J.~M.}\ \bibnamefont
  {Horowitz}}\ and\ \bibinfo {author} {\bibfnamefont {M.}~\bibnamefont
  {Esposito}},\ }\href@noop {} {\bibfield  {journal} {\bibinfo  {journal}
  {Physical Review X}\ }\textbf {\bibinfo {volume} {4}},\ \bibinfo {pages}
  {031015} (\bibinfo {year} {2014})}\BibitemShut {NoStop}%
\bibitem [{\citenamefont {Nicoletti}\ and\ \citenamefont
  {Busiello}(2021)}]{nicoletti2021mutual}%
  \BibitemOpen
  \bibfield  {author} {\bibinfo {author} {\bibfnamefont {G.}~\bibnamefont
  {Nicoletti}}\ and\ \bibinfo {author} {\bibfnamefont {D.~M.}\ \bibnamefont
  {Busiello}},\ }\href@noop {} {\bibfield  {journal} {\bibinfo  {journal}
  {Physical review letters}\ }\textbf {\bibinfo {volume} {127}},\ \bibinfo
  {pages} {228301} (\bibinfo {year} {2021})}\BibitemShut {NoStop}%
\bibitem [{\citenamefont {Tka{\v{c}}ik}\ and\ \citenamefont
  {Bialek}(2016)}]{tkavcik2016information}%
  \BibitemOpen
  \bibfield  {author} {\bibinfo {author} {\bibfnamefont {G.}~\bibnamefont
  {Tka{\v{c}}ik}}\ and\ \bibinfo {author} {\bibfnamefont {W.}~\bibnamefont
  {Bialek}},\ }\href@noop {} {\bibfield  {journal} {\bibinfo  {journal} {Annual
  Review of Condensed Matter Physics}\ }\textbf {\bibinfo {volume} {7}},\
  \bibinfo {pages} {89} (\bibinfo {year} {2016})}\BibitemShut {NoStop}%
\bibitem [{\citenamefont {Gillespie}(1977)}]{gillespie1977exact}%
  \BibitemOpen
  \bibfield  {author} {\bibinfo {author} {\bibfnamefont {D.~T.}\ \bibnamefont
  {Gillespie}},\ }\href@noop {} {\bibfield  {journal} {\bibinfo  {journal} {The
  journal of physical chemistry}\ }\textbf {\bibinfo {volume} {81}},\ \bibinfo
  {pages} {2340} (\bibinfo {year} {1977})}\BibitemShut {NoStop}%
\bibitem [{\citenamefont {Esposito}(2012)}]{espositoCG}%
  \BibitemOpen
  \bibfield  {author} {\bibinfo {author} {\bibfnamefont {M.}~\bibnamefont
  {Esposito}},\ }\href@noop {} {\bibfield  {journal} {\bibinfo  {journal}
  {Physical Review E}\ }\textbf {\bibinfo {volume} {85}},\ \bibinfo {pages}
  {041125} (\bibinfo {year} {2012})}\BibitemShut {NoStop}%
\bibitem [{\citenamefont {Busiello}\ \emph {et~al.}(2019)\citenamefont
  {Busiello}, \citenamefont {Hidalgo},\ and\ \citenamefont
  {Maritan}}]{busielloNJP}%
  \BibitemOpen
  \bibfield  {author} {\bibinfo {author} {\bibfnamefont {D.~M.}\ \bibnamefont
  {Busiello}}, \bibinfo {author} {\bibfnamefont {J.}~\bibnamefont {Hidalgo}}, \
  and\ \bibinfo {author} {\bibfnamefont {A.}~\bibnamefont {Maritan}},\
  }\href@noop {} {\bibfield  {journal} {\bibinfo  {journal} {New Journal of
  Physics}\ }\textbf {\bibinfo {volume} {21}},\ \bibinfo {pages} {073004}
  (\bibinfo {year} {2019})}\BibitemShut {NoStop}%
\bibitem [{\citenamefont {Busiello}\ and\ \citenamefont
  {Maritan}(2019)}]{busielloJSTAT}%
  \BibitemOpen
  \bibfield  {author} {\bibinfo {author} {\bibfnamefont {D.~M.}\ \bibnamefont
  {Busiello}}\ and\ \bibinfo {author} {\bibfnamefont {A.}~\bibnamefont
  {Maritan}},\ }\href@noop {} {\bibfield  {journal} {\bibinfo  {journal}
  {Journal of Statistical Mechanics: Theory and Experiment}\ }\textbf {\bibinfo
  {volume} {2019}},\ \bibinfo {pages} {104013} (\bibinfo {year}
  {2019})}\BibitemShut {NoStop}%
\bibitem [{\citenamefont {Callen}(1998)}]{callen}%
  \BibitemOpen
  \bibfield  {author} {\bibinfo {author} {\bibfnamefont {H.~B.}\ \bibnamefont
  {Callen}},\ }\href@noop {} {\enquote {\bibinfo {title} {Thermodynamics and an
  introduction to thermostatistics},}\ } (\bibinfo {year} {1998})\BibitemShut
  {NoStop}%
\bibitem [{\citenamefont {Proesmans}\ \emph
  {et~al.}(2016{\natexlab{c}})\citenamefont {Proesmans}, \citenamefont
  {Cleuren},\ and\ \citenamefont {Van~den Broeck}}]{PhysRevLett.116.220601}%
  \BibitemOpen
  \bibfield  {author} {\bibinfo {author} {\bibfnamefont {K.}~\bibnamefont
  {Proesmans}}, \bibinfo {author} {\bibfnamefont {B.}~\bibnamefont {Cleuren}},
  \ and\ \bibinfo {author} {\bibfnamefont {C.}~\bibnamefont {Van~den Broeck}},\
  }\href {\doibase 10.1103/PhysRevLett.116.220601} {\bibfield  {journal}
  {\bibinfo  {journal} {Phys. Rev. Lett.}\ }\textbf {\bibinfo {volume} {116}},\
  \bibinfo {pages} {220601} (\bibinfo {year} {2016}{\natexlab{c}})}\BibitemShut
  {NoStop}%
\bibitem [{\citenamefont {Proesmans}\ and\ \citenamefont {Van~den
  Broeck}(2015)}]{proesmans2015onsager}%
  \BibitemOpen
  \bibfield  {author} {\bibinfo {author} {\bibfnamefont {K.}~\bibnamefont
  {Proesmans}}\ and\ \bibinfo {author} {\bibfnamefont {C.}~\bibnamefont
  {Van~den Broeck}},\ }\href
  {https://journals.aps.org/prl/abstract/10.1103/PhysRevLett.115.090601}
  {\bibfield  {journal} {\bibinfo  {journal} {Physical review letters}\
  }\textbf {\bibinfo {volume} {115}},\ \bibinfo {pages} {090601} (\bibinfo
  {year} {2015})}\BibitemShut {NoStop}%
\bibitem [{\citenamefont {Proesmans}\ \emph
  {et~al.}(2016{\natexlab{d}})\citenamefont {Proesmans}, \citenamefont
  {Dreher}, \citenamefont {Gavrilov}, \citenamefont {Bechhoefer},\ and\
  \citenamefont {Van~den Broeck}}]{proesmans16b}%
  \BibitemOpen
  \bibfield  {author} {\bibinfo {author} {\bibfnamefont {K.}~\bibnamefont
  {Proesmans}}, \bibinfo {author} {\bibfnamefont {Y.}~\bibnamefont {Dreher}},
  \bibinfo {author} {\bibfnamefont {M.~c.~v.}\ \bibnamefont {Gavrilov}},
  \bibinfo {author} {\bibfnamefont {J.}~\bibnamefont {Bechhoefer}}, \ and\
  \bibinfo {author} {\bibfnamefont {C.}~\bibnamefont {Van~den Broeck}},\ }\href
  {\doibase 10.1103/PhysRevX.6.041010} {\bibfield  {journal} {\bibinfo
  {journal} {Phys. Rev. X}\ }\textbf {\bibinfo {volume} {6}},\ \bibinfo {pages}
  {041010} (\bibinfo {year} {2016}{\natexlab{d}})}\BibitemShut {NoStop}%
\bibitem [{\citenamefont {Proesmans}\ and\ \citenamefont
  {Fiore}(2019)}]{fiorek}%
  \BibitemOpen
  \bibfield  {author} {\bibinfo {author} {\bibfnamefont {K.}~\bibnamefont
  {Proesmans}}\ and\ \bibinfo {author} {\bibfnamefont {C.~E.}\ \bibnamefont
  {Fiore}},\ }\href@noop {} {\bibfield  {journal} {\bibinfo  {journal}
  {Physical Review E}\ }\textbf {\bibinfo {volume} {100}},\ \bibinfo {pages}
  {022141} (\bibinfo {year} {2019})}\BibitemShut {NoStop}%
\bibitem [{\citenamefont {Kedem}\ and\ \citenamefont
  {Caplan}(1965)}]{TF9656101897}%
  \BibitemOpen
  \bibfield  {author} {\bibinfo {author} {\bibfnamefont {O.}~\bibnamefont
  {Kedem}}\ and\ \bibinfo {author} {\bibfnamefont {S.~R.}\ \bibnamefont
  {Caplan}},\ }\href {\doibase 10.1039/TF9656101897} {\bibfield  {journal}
  {\bibinfo  {journal} {Trans. Faraday Soc.}\ }\textbf {\bibinfo {volume}
  {61}},\ \bibinfo {pages} {1897} (\bibinfo {year} {1965})}\BibitemShut
  {NoStop}%
\bibitem [{\citenamefont {Fiore}\ and\ \citenamefont
  {da~Luz}(2013)}]{fiorejcp2013}%
  \BibitemOpen
  \bibfield  {author} {\bibinfo {author} {\bibfnamefont {C.~E.}\ \bibnamefont
  {Fiore}}\ and\ \bibinfo {author} {\bibfnamefont {M.~G.~E.}\ \bibnamefont
  {da~Luz}},\ }\href {\doibase 10.1063/1.4772809} {\bibfield  {journal}
  {\bibinfo  {journal} {The Journal of Chemical Physics}\ }\textbf {\bibinfo
  {volume} {138}},\ \bibinfo {pages} {014105} (\bibinfo {year} {2013})},\
  \Eprint {http://arxiv.org/abs/https://doi.org/10.1063/1.4772809}
  {https://doi.org/10.1063/1.4772809} \BibitemShut {NoStop}%
\bibitem [{\citenamefont {Challa}\ \emph {et~al.}(1986)\citenamefont {Challa},
  \citenamefont {Landau},\ and\ \citenamefont {Binder}}]{challa}%
  \BibitemOpen
  \bibfield  {author} {\bibinfo {author} {\bibfnamefont {M.~S.~S.}\
  \bibnamefont {Challa}}, \bibinfo {author} {\bibfnamefont {D.~P.}\
  \bibnamefont {Landau}}, \ and\ \bibinfo {author} {\bibfnamefont
  {K.}~\bibnamefont {Binder}},\ }\href {\doibase 10.1103/PhysRevB.34.1841}
  {\bibfield  {journal} {\bibinfo  {journal} {Phys. Rev. B}\ }\textbf {\bibinfo
  {volume} {34}},\ \bibinfo {pages} {1841} (\bibinfo {year}
  {1986})}\BibitemShut {NoStop}%
\end{thebibliography}%

\newpage
\clearpage

\appendix            

\onecolumngrid
\begin{center}
\textbf{\large Supplemental Material: Powerful ordered collective heat engines}
\end{center}

\begin{center}
Fernando S. Filho, Gustavo A. L. Forão, Daniel M. Busiello, Bart Cleuren and C. E. Fiore
\end{center}

This supplemental material is structured as follows: In Sec.~\ref{aptr} we describe the transition rates for finite $N$ and $N\rightarrow \infty$ interacting units. In Sec.~\ref{apc}, we derive the effective model for steady probabilities in the regime of strong collective effects as a function of $\beta_1$, $\beta_2$, $\epsilon$ and $F$. The main results for $q=2$ engines and the analysis of a minimal setup composed of $N=2$ ($q=3$) interacting engines are shown in Secs.~\ref{apt} and \ref{apn2}, respectively. System features near equilibrium are investigated in Sec.~\ref{lin} through a linear analysis. The crossover from collective to independent regimes is described in Sec.~\ref{pump}. As additional investigations, the linear stability of the disordered solution for $q=3$ and a comparison between the all-to-all case and local interactions are investigated     in 
 Secs.~\ref{apb} and \ref{aph}, respectively.

\section{Transition rates}\label{aptr}
As stated in the main text, collective effects from ordered structures have been investigated for two models (A and B), for $q=2$ and $q=3$. When $q=2$, model B can be derived from model A setting $\alpha=0$. The system dynamics is governed by the master equation ${\dot p}_j=\sum_{\nu=1}^2\sum_{i\neq j}(\omega^{(\nu)}_{ji}p_i-\omega^{(\nu)}_{ij}p_j) $, where the transition rates from $i$ to $j$ are given by
\begin{equation}
\omega^{(1)}_{ji}=\Gamma e^{-\frac{\beta_1}{2}\{\mp\epsilon(1+\alpha)(1-\frac{2N_{i\uparrow(\downarrow)}\mp 1}{N}) \mp F\}} \quad {\rm and} \quad
\omega^{(2)}_{ji}=\Gamma e^{-\frac{\beta_2}{2}\{\mp\epsilon(1+\alpha)(1-\frac{2N_{i\uparrow(\downarrow)} \mp 1}{N}) \pm F\}},
\end{equation}
where the sign of $\mp\epsilon(1+\alpha)$ accounts for the fact that $N^{(j)}_{\uparrow} = N^{(i)}_{\uparrow} \pm 1$ and $N^{(i)}_{\downarrow} = N_{i\downarrow} \mp 1$. Analogously, the sign of the driving depends on initial and final states and on the bath to which the system is coupled, i.e., $\mp F$ for the cold bath ($\nu = 1$) and $\pm F$ for the hot bath ($\nu=2$). If $N$ is finite, the dynamics can be simulated using a standard Gillespie algorithm \cite{gillespie1977exact}.
  
In the thermodynamic limit $ N\rightarrow \infty$, we can employ a mean-field approach. We introduce the mean occupation density of a given state, $p_{\uparrow(\downarrow)} = \langle \sum_i N^{(i)}_{\uparrow(\downarrow)}/N\rangle$, which are characterized only by the index of states, massively reducing the complexity of the equations. By employing the mean-field approximation of writing down any $n$-point correlations as the product of $n$ averages, $p_{\uparrow(\downarrow)}$ is ruled by the master equation ${\dot p}_\beta=\sum_{\nu=1}^2J_{\beta \beta'}^{(\nu)}$, where $J_{\beta \beta'}^{(\nu)}=\omega^{(\nu)}_{\beta \beta'}p_{\beta'}-\omega^{(\nu)}_{\beta'\beta}p_\beta$ with transition rates listed below:
\begin{eqnarray}
\omega^{(1)}_{\uparrow\downarrow}=\Gamma e^{-\frac{\beta_1}{2}\{-\epsilon(1+\alpha)(1-2p_{\uparrow}) - F\}} \quad {\rm and} \quad
\omega^{(2)}_{\uparrow\downarrow}=\Gamma e^{-\frac{\beta_2}{2}\{-\epsilon(1+\alpha)(1-2p_{\uparrow}) + F\}} \;, \\
\omega^{(1)}_{\downarrow\uparrow}=\Gamma e^{-\frac{\beta_1}{2}\{\epsilon(1+\alpha)(1-2p_{\uparrow}) + F\}}\quad {\rm and} \quad
\omega^{(2)}_{\downarrow\uparrow}=\Gamma e^{-\frac{\beta_2}{2}\{\epsilon(1+\alpha)(1-2p_{\uparrow}) - F\}} \;.
\end{eqnarray}

Transition rates are evaluated in a similar fashion for $q=3$. Starting with model A, they are identical to $q=2$ for transitions of type $\uparrow\rightarrow \downarrow$ and $\downarrow\rightarrow \uparrow$, whereas the energy difference reads $\epsilon(N^{(i)}_k-\alpha N^{(i)}_\ell)/N$ for transitions like $0\rightarrow \uparrow(\downarrow)$, where $k=\uparrow(\downarrow)$ and $\ell= \downarrow(\uparrow)$. All the remaining ones can be analogously computed. Likewise, for model B, a given transition $N^{(j)}_{\ell} = N^{(i)}_{\ell}- 1$ and $N^{(i)}_{k} = N^{(i)}_{k}+ 1$ (where $k$, $\ell \in (\uparrow,0,\downarrow)$) has energy difference given by $\epsilon (N^{(i)}_k-N^{(i)}_\ell+1)/N$ \cite{herpich,herpich2}. Numerical simulations are performed as before, but now there are $2\,q\,(q-1) = 12$ distinct transitions. As for $q=2$, the limit $N\rightarrow \infty$ is promptly obtained and described by the master equation ${\dot p}_\beta=\sum_{\nu=1}^2\sum_{\beta'\neq \beta} J_{\beta \beta'}^{(\nu)}$ [$\beta\in(\downarrow,0,\uparrow)$]. For model A, some of the transition rates are:
\begin{equation}
\omega^{(1)}_{\uparrow\downarrow}=\Gamma e^{-\frac{\beta_1}{2}\{-\epsilon(1+\alpha)(p_\downarrow-p_\uparrow) + F\}}, \qquad
\omega^{(1)}_{\uparrow 0}=\Gamma e^{-\frac{\beta_1}{2}\{\epsilon(p_\uparrow-\alpha p_\downarrow)- F\}} \quad {\rm and} \quad
\omega^{(1)}_{0\downarrow}=\Gamma e^{-\frac{\beta_1}{2}\{\epsilon(\alpha p_\uparrow- p_\downarrow)- F\}} \;.
\end{equation}
For model B, we have:
\begin{equation}
\omega^{(1)}_{\uparrow\downarrow}=\Gamma e^{-\frac{\beta_1}{2}\{\epsilon(p_\uparrow-p_\downarrow) + F\}}, \qquad
\omega^{(1)}_{\uparrow 0}=\Gamma e^{-\frac{\beta_1}{2}\{\epsilon(p_\uparrow-p_0)- F\}}
\quad {\rm and} \quad
\omega^{(1)}_{0\downarrow}=\Gamma e^{-\frac{\beta_1}{2}\{\epsilon(p_0- p_\downarrow)- F\}} \;
\end{equation}
where all the others can be easily computed along the same line, remembering also that $\omega^{(2)}_{ij}$  is promptly obtained from $\omega^{(1)}_{ij}$ just by replacing $F\rightarrow -F$. The thermodynamic quantities in the limit $N\rightarrow \infty$ are similar in form to those presented in the main text (for finite $N$). Indeed, the power ${\cal P}$ and heat $\langle\dot{Q}_\nu\rangle$ per unit are given by
\begin{equation}
\label{work1}
{\cal P}=-\sum_{(\nu,\gamma)} F_\gamma^{(\nu)}\sum_{(\beta,\beta')}d_{\beta \beta'}^{(\nu)}J_{\beta \beta'}^{(\nu)} \qquad{\rm and} \qquad
\langle \dot{Q}_\nu\rangle=\sum_{(\beta,\beta')}\left(E_\beta-E_{\beta'}+\sum_{\gamma}F_\gamma^{(\nu)}d_{\beta \beta'}^{(\nu)}\right)J_{\beta \beta'}^{(\nu)}, 
\end{equation}
with the energy difference $E_\beta-E_{\beta'}$ is the same quantity appearing also in the exponent of the transition rates.

\section{Effective description for the probability distribution in the regime of strong collective effects} \label{apc}

\begin{figure}[t]
\includegraphics[scale=0.5]{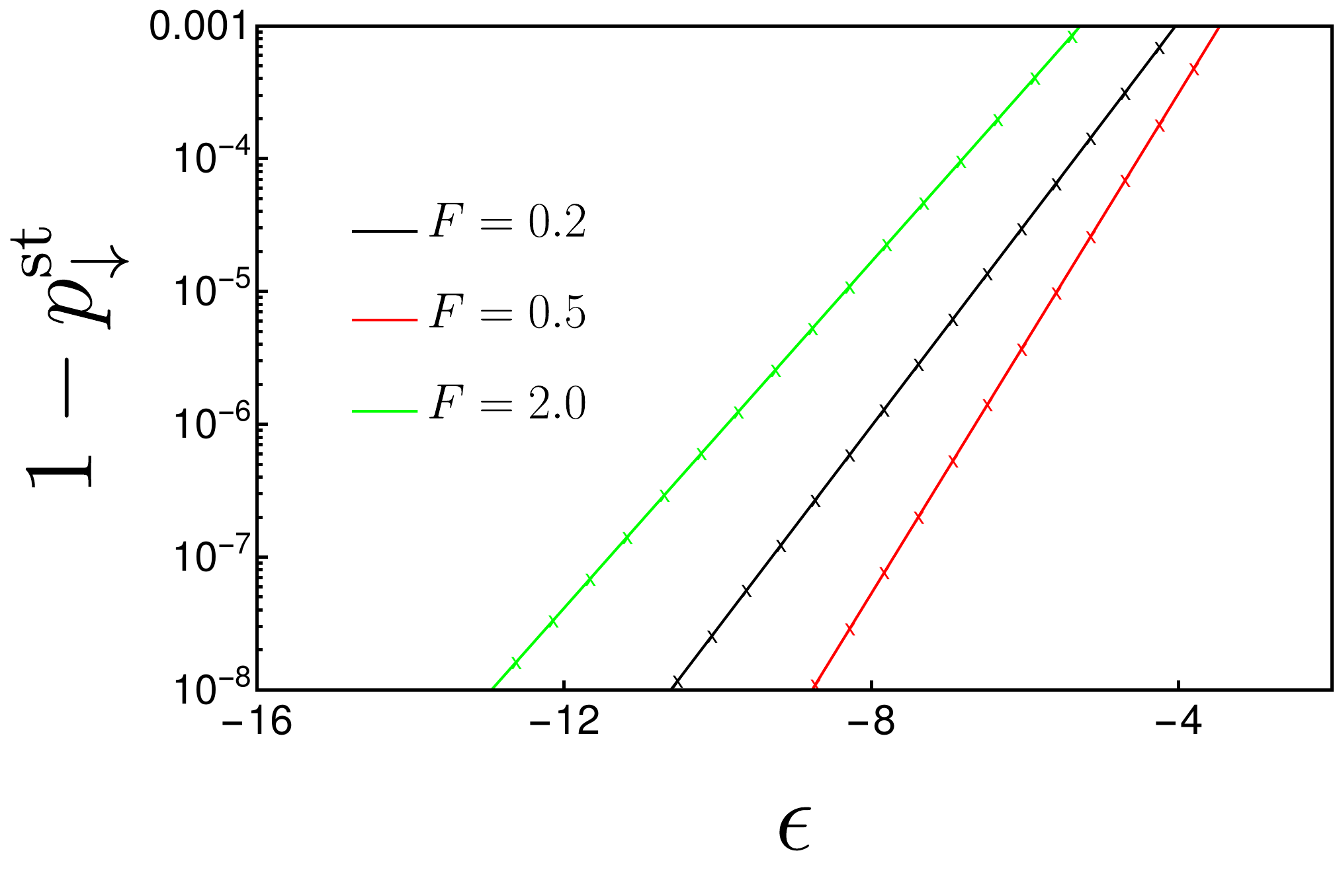} 
\caption{Model A, $q=3$. Semilog plot of $1-p^{st}_\downarrow$ versus $\epsilon$ for distinct sets of $\beta_1$ and $F$. Continuous lines are exact results for $N\rightarrow \infty$, while symbols corresponds to the solution evaluated from the two-state effective model. Black, red and green curves show results for $\beta_1=5/2,10/3$ and $2$, respectively. In all cases $\beta_2=1$. }
\label{appr}
\end{figure}

Despite the nonlinear shape of the master equation for $N\rightarrow \infty$, it is possible to get some insights about the probability distribution in the regime of strong collective effects. Starting with $q=2$, the  ordered phase is two-fold degenerate and characterized by the predominance of spins of one type ($p^{st}_\downarrow$ or $p^{st}_\uparrow$). By focusing on the case $p^{st}_\downarrow \approx 1$, its steady-state probability is 
\begin{equation}
p^{st}_\downarrow \approx \frac{\omega^{(1)}_{\downarrow\uparrow}+\omega^{(2)}_{\downarrow\uparrow}}{\omega^{(1)}_{\downarrow\uparrow}+\omega^{(2)}_{\downarrow\uparrow}+\omega^{(1)}_{\uparrow\downarrow}+\omega^{(2)}_{\uparrow\downarrow}}\;.
\label{theor}
\end{equation}
This is an implicit equation, as transition rates depend on $p_\downarrow$. Inserting the expression of the transition rates derived in Sec.~\ref{aptr}, and performing the $\epsilon \to -\infty$ limit, we have:
\begin{equation}
p^{st}_\downarrow \approx \frac{1}{1+e^{\frac{1}{2} \{(\beta_1+\beta_2)(\alpha +1) \epsilon  +F (\beta_1-\beta_2)\}}}.
\label{theor1}
\end{equation}
Taking into account that $\beta_2<\beta_1$, $F>0$ and $-\epsilon \gg F$, as we are in the regime of strong collective effects, we can approximate the above expression as $p^{st}_{\downarrow}\approx\frac{1}{2}e^{\{(\beta_1+\beta_2)(1+\alpha)\epsilon\}}$$e^{\frac{(\beta_1-\beta_2)F}{2}}$. This can also be derived from the fact that $p^{st}_{\downarrow}\approx 1 - (\omega^{(1)}_{\uparrow\downarrow} + \omega^{(2)}_{\uparrow\downarrow})/(\omega^{(1)}_{\downarrow\uparrow} + \omega^{(2)}_{\downarrow\uparrow}) \approx 1 - \omega^{(1)}_{\uparrow\downarrow}/\omega^{(2)}_{\downarrow\uparrow}$, under the aforementioned assumptions.
By inserting this expression for $p^{st}_\downarrow$ into Eq.~\eqref{work1}, and considering that $p^{st}_\uparrow = 1 - p^{st}_\downarrow$, one arrives at the expressions for ${\cal P}_{\rm eff}$ per unit when $q=2$:
\begin{eqnarray}
\label{twork1}
{\cal P}_{\rm eff} &=& \frac{F}{2}  e^{-\frac{1}{2} \beta_2[(\alpha +1) \epsilon M+F]}\Big[e^{\frac{1}{2}[ (\beta_1+\beta_2)F-(\beta_1-\beta_2)(\alpha +1) \epsilon  M]}-1\Big] \Big[(1+M) e^{\frac{1}{2}[-(\beta_1-\beta_2)F+(\beta_1+\beta_2)(\alpha +1) \epsilon M ]}-M+1\Big] \;. \nonumber
\end{eqnarray}
and the following for 
$\langle \dot{Q}_2\rangle_{\rm eff}$:
\begin{equation}
\langle \dot{Q}_2\rangle_{\rm eff}=   -(F+(\alpha +1) M \epsilon )\Big[\sinh \left(\frac{\beta_2}{2} (F+(\alpha +1) M \epsilon )\right)+M \cosh \left(\frac{\beta_2}{2} (F+(\alpha +1) M \epsilon )\right)\Big].
\label{theat0}
\end{equation}
with $M = p_\downarrow^{\rm st} - p_\uparrow^{\rm st}$. It is worth mentioning that $|M|$ reduces to $1 - 2e^{\beta (1+\alpha)\epsilon}$ in the equilibrium regime ($\beta_1 = \beta_2$ and $F = 0$), becoming equal to the magnetization per spin of the Ising model for sufficiently low temperatures $\beta \gg \beta_c = (1+\alpha)\epsilon/k_B$. Notice that we used a different notation with respect to Eqs.~\eqref{work} and \eqref{heat} since, in general, these quantities might be different due to the coarse-graining procedure \cite{espositoCG,busielloNJP,busielloJSTAT}.
When $q=3$, $p_\uparrow \approx 0$, hence the state $\uparrow$ can be seen as a source, meaning that, at stationarity, $p_\downarrow$ and $p_0$ sum up to $1$ and satisfy detailed balance. As a consequence, the system can be seen as a $2$-state system composed by states $0$ and $\downarrow$ only. Hence, retracing the procedure described above, we have:
\begin{equation}
p^{st}_{\downarrow} \approx \frac{1}{1+e^{\frac{1}{2} \{ (\beta_1+\beta_2)\epsilon + (\beta_1-\beta_2)F\}}} \;,
\end{equation}
with $p^{st}_{0}\approx 1-p^{st}_{\downarrow}$. Once again, as before, we can also write $p^{st}_\downarrow \approx 1 -\omega^{(1)}_{0\downarrow}/\omega^{(2)}_{\downarrow0}=1-e^{\frac{1}{2}\{(\beta_1+\beta_2)\epsilon+(\beta_1-\beta_2)F\}}$, which gives our approximation for strong collective effects. By inserting this expression into Eq.~\eqref{work1}, we obtain:
\begin{equation}
\langle \dot{Q}_2\rangle_{\rm eff}=-(1+M) \Big[(F+M \epsilon ) e^{\frac{1}{2} \beta_2 (F+M \epsilon )}-(F+\alpha  M \epsilon ) e^{-\frac{1}{2} \beta_2 (F+\alpha  M \epsilon)}\Big]+M \Big[(F-(\alpha +1) M \epsilon ) e^{\frac{1}{2} \beta_2 (F-(\alpha +1) M\epsilon)}-(F+M \epsilon ) e^{-\frac{1}{2} \beta_2 (F+M \epsilon )}\Big].
\label{aph2}
\end{equation}
Fig. \ref{appr} shows the validity of our approximate expressions for $p^{st}_{\downarrow}$ for distinct sets of temperatures 
$\beta_1,\beta_2$ and $F$, when $q=3$.
    
\section{Thermodynamics of q=2 engines}\label{apt}
\begin{figure}[h]
\centering
\includegraphics[scale=0.23]{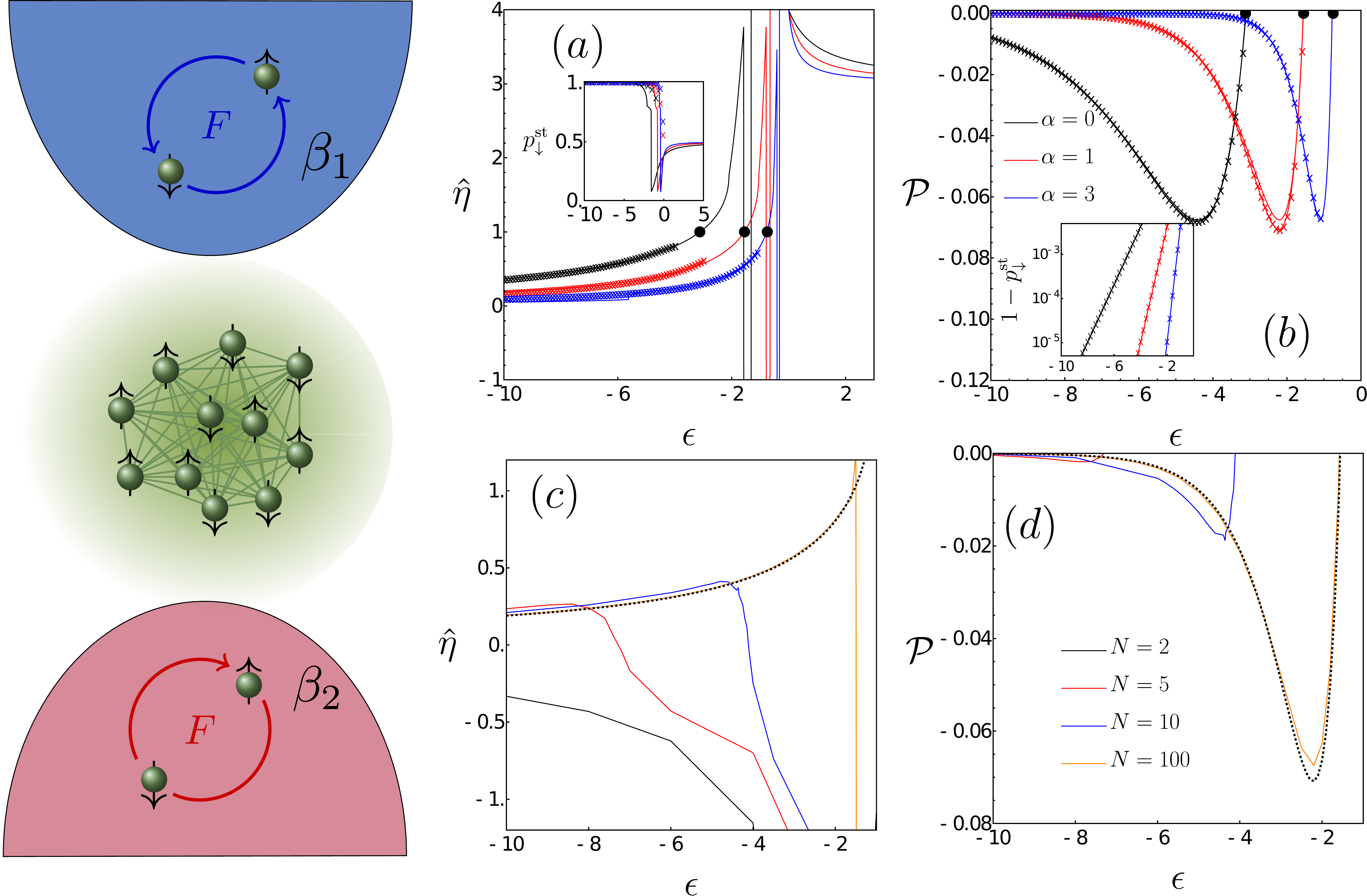}
\caption{Left: Schematics of a $q=2$ engine. For $\beta_1=2,\beta_2=F=1$ and distinct $\alpha$, panels (a) and (b) depict the efficiency ${\hat \eta}$ and the power per unit ${\cal P}$ versus the interaction strength $\epsilon$, respectively. Circles show the optimal efficiency, $\eta_{ME}=\eta_c$ (${\hat \eta}_{ME}=1$) in this setting. Symbols correspond to the effective two-state description. Insets: plot of $p^{st}_\downarrow$ (panel (a)) and semilog plot $p_\uparrow$ (panel (b)) versus $\epsilon$. For the same parameters, panels (c) and (d) show ${\cal P} = \langle{\mathtt P}\rangle/N$ and ${\hat \eta}$ for different (finite) $N$ and $\alpha=1$.}
\label{fig0}
\end{figure}

\begin{figure}[h]
\includegraphics[scale=0.35]{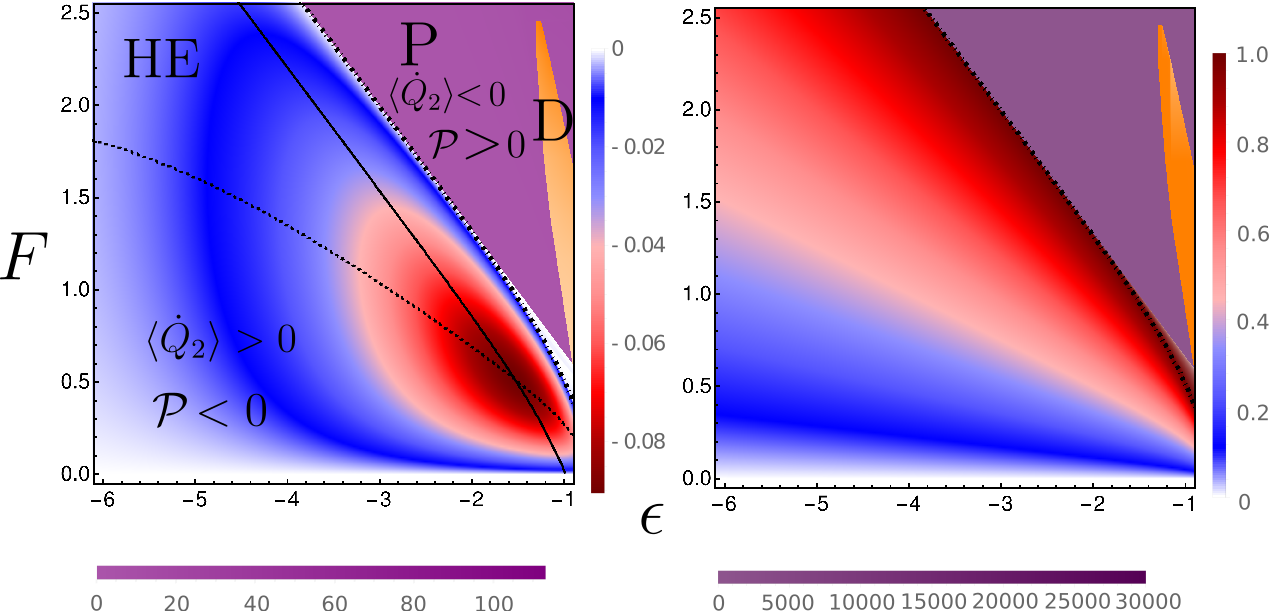} 
\caption{Heat maps for ${\cal P}$ and ${\hat \eta}$ for the same parameters as in Fig.~\ref{fig0}. Heat engine, pump, and dud regimes are described by symbols HE, P, and D, respectively. Continuous and dotted lines denote the maximization of power respectively holding $F$ and $\epsilon$ fixed. The dot-dashed line corresponds to the crossover from heat engine to pump regimes, in which $\eta=\eta_c$ in this setup ($q=2$).}
\label{fig4}
\end{figure}

The main features of the thermodynamics of $q=2$ engines are summarized in Figs.~\ref{fig0} and \ref{fig4}. Fig.~\ref{fig0} shows, for finite $N$ and $N\rightarrow \infty$, power per unit, efficiency and reliability of the two-state effective model discussed above.
Fig.~\ref{fig4} extends the power and efficiency heat maps to the case $q=2$ (and $\alpha=1$),
 showing that the system exhibits a very similar behavior with respect to $q=3$ (main text). Furthermore, Carnot efficiency is reached for $q=2$ for all values of $\alpha$, such that ${\cal P}=\langle {\dot Q_2}\rangle =0$ and $M$ satisfies the implicit equation $M=-\tanh \left[\frac{\beta_2}{2} (F+(\alpha +1) M \epsilon )\right]$, obtained from the effective two-state description.

\section{Heat maps for q=3 and N=2 engines}\label{apn2}
\begin{figure}[h]
\centering
\includegraphics[scale=0.32]{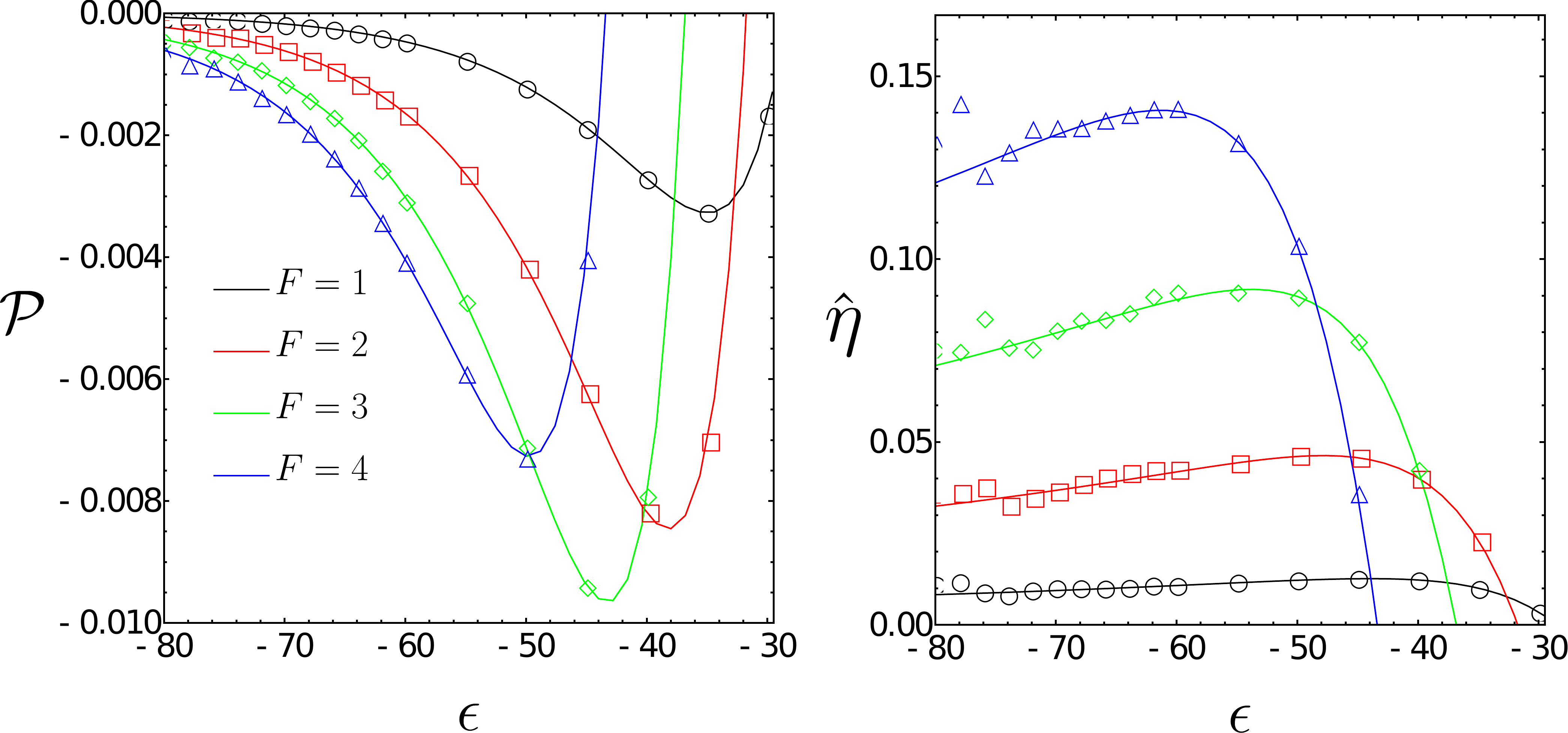}
\caption{Model A, $q=3$. Left and right panels show the power per unit ${\cal P}=\langle{\mathtt P}\rangle/N$ and efficiency ${\hat \eta}$ for $N=2$ for different values of $F$. Continuous lines are exact solutions obtained from the microscopic master equation, while symbols come from numerical simulations using the Gillespie algorithm. Parameters: $\alpha=1$, $\beta_1=1$, and $\beta_2=0.4$.}
\label{n2exactvsgill}
\end{figure}

\begin{figure}[h]
\includegraphics[scale=0.35]{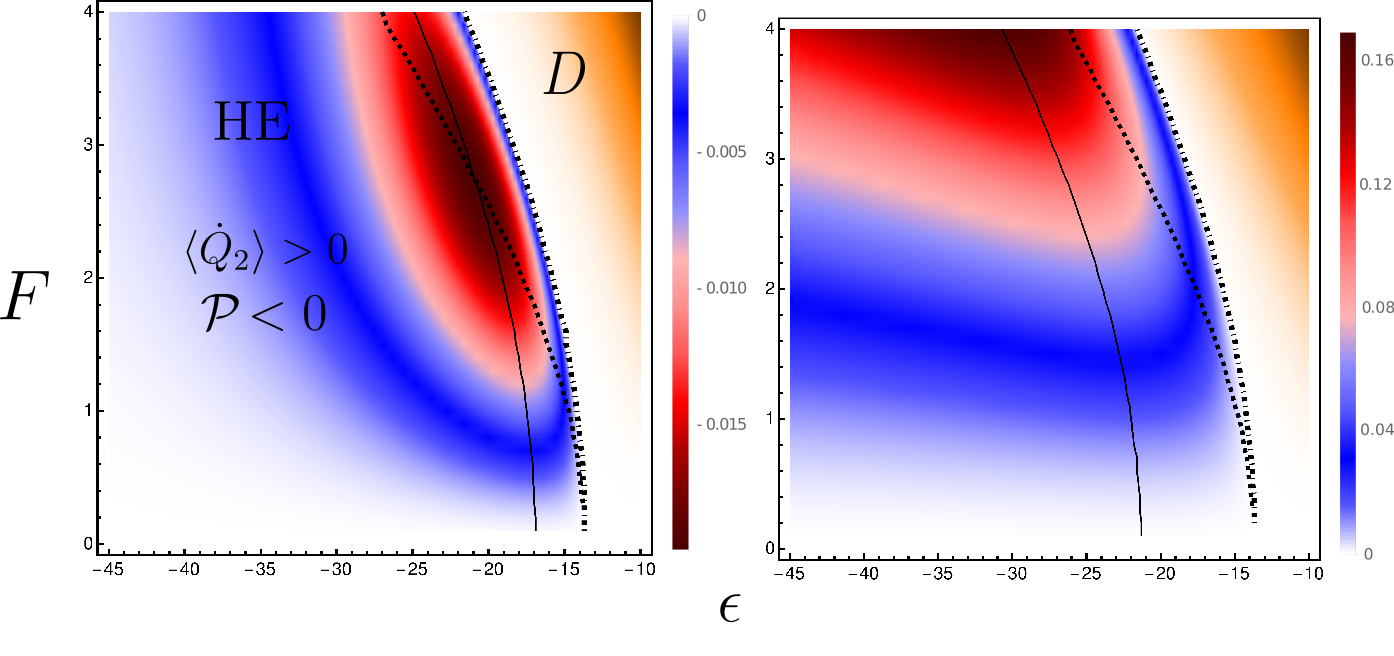}
\caption{Model A, $q=3$ and $N=2$. From left to right, top panels depict power per unit ${\cal P} = \langle{\mathtt P}\rangle/N$ and efficiency ${\hat \eta}$ heat maps. Heat engine and dud regimes are denoted by HE and D, respectively. For a better visualization, the pump regime has not been indicated. Parameters: $\alpha=1$, $\beta_1=1$ and $\beta_2=0.4$.}
\label{fig5n}
\end{figure}

This section discusses two important aspects introduced in the main text: the reliability of numerical simulations for finite $N$ and the fact that a minimal setup of $N=2$ interacting units already captures the essential ingredients of the model. Results are shown for $\beta_1=1$ and $\beta_2=0.4$ only for the sake of a better visualization. Fig.~\ref{n2exactvsgill} compares thermodynamic quantities evaluated from numerical simulations (Gillespie algorithm) and those from exact steady probabilities computed from the microscopic master equation for $N=2$. Fig.~\ref{fig5n} extends the heat maps to $N=2$, showing that despite the substantial reduced performance, all characteristics from collective effects are already present in this minimal setup.

\section{Linear regime}\label{lin}
\begin{figure}[h]
\centering
\includegraphics[scale=0.3]{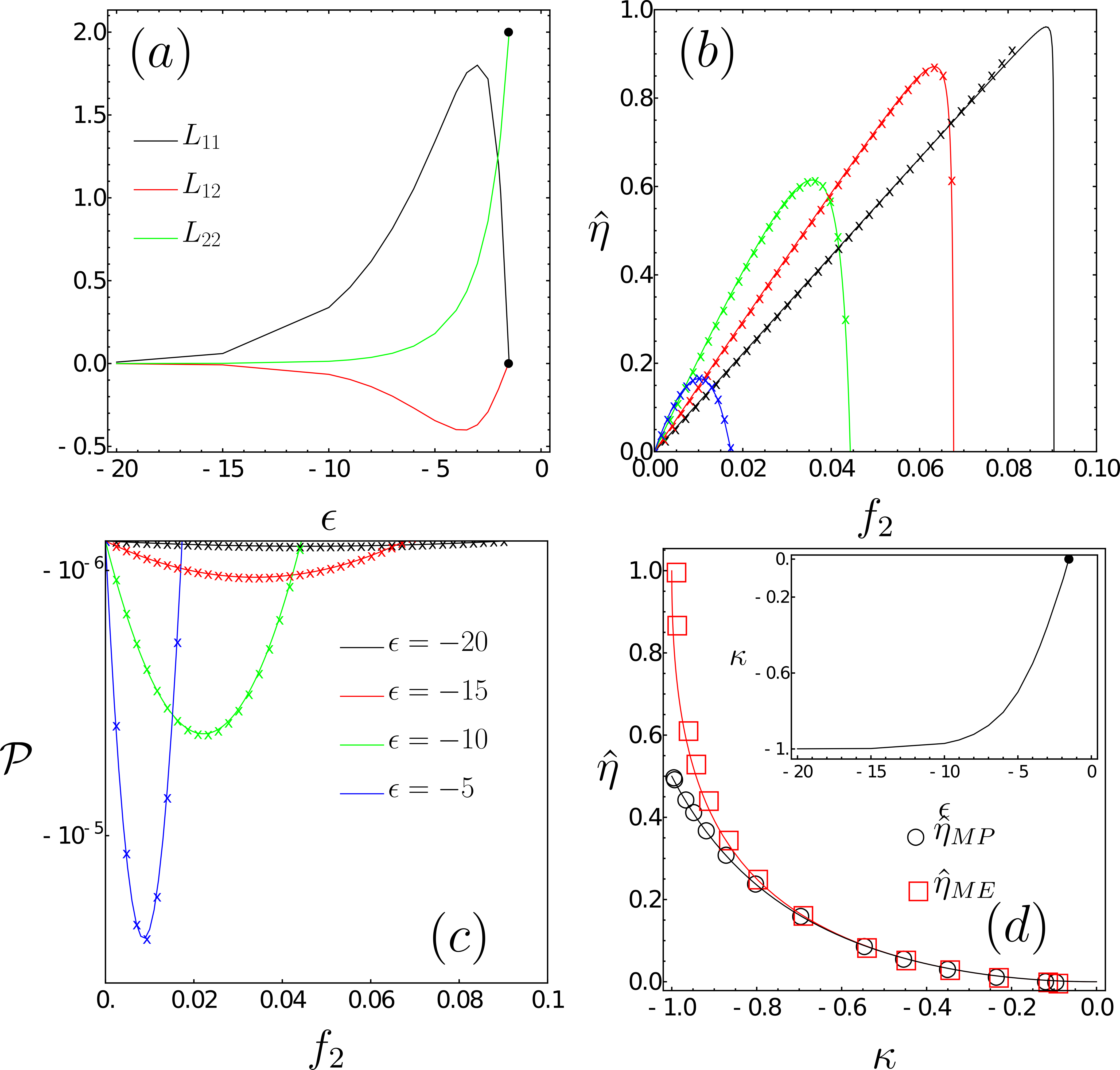} 
\caption{For model A, $q=3$ and $\alpha=1$, we show the thermodynamics of the system close to the equilibrium regime. Panel (a) show the Onsager coefficients $L_{ij}$ versus the interaction parameter $\epsilon$. In (b) and (c), respectively $\hat{\eta}$ and ${\cal P}$ versus $f_2$ are reported for $f_1=9.10^{-3}$. Continuous lines are exact results, while symbols correspond to Eq.~\eqref{efff}. The heat engine behavior is delimited by vertical lines at $f_m$. Panel (d) shows the behavior of maximum efficiency $\hat{\eta}_{\maxe}$ and efficiency at maximum power $\hat{\eta}_{\maxp}$ versus $\kappa$, where continuous lines follow Eqs.~\ref{etame1} and \eqref{etamp1}. The inset show how $\kappa$ changes as a function of $\epsilon$, and $\bullet$ denotes the phase transition
to the independent regime taking place at $\epsilon_c \approx -1.5$ (see also the same symbol in panel (a).}
\label{ons}
\end{figure}

Additional insights regarding the influence of the collective effects on the efficiency and power of our system can be obtained from a linear analysis, valid near equilibrium ($\beta_1-\beta_2$, $F<<1$). By resorting to the ideas of linear stochastic thermodynamics \cite{callen,PhysRevLett.116.220601,proesmans2015onsager,proesmans16b,fiorek}, we introduce the following thermodynamic forces  $f_1=\beta_1-\beta_2$ and $f_2=\beta_1F$, in such a way that the entropy production, $\langle {\dot \sigma} \rangle$, is expressed in the bilinear form
\begin{equation}
\langle {\dot \sigma} \rangle =\left(\beta_1 - \beta_2\right) \langle \dot{Q}_2\rangle+\beta_1\mathcal{P}= J_1f_1+J_2f_2,
\end{equation}
where $J_1=\langle \dot{Q}_2\rangle$ and $J_2=\mathcal{P}/F$ denote the thermodynamic fluxes. Close to equilibrium, i.e., in the linear regime, these fluxes can be expressed in terms of the Onsager coefficients, $J_1=L_{11}f_1+L_{12}f_2$ and  $J_2=L_{21}f_1+L_{22}f_2$, which satisfy the conditions $L_{11}$, $L_{22} \ge 0$ and $L_{12}=L_{21}$. From the equation above, the efficiency $\hat{\eta}$ promptly reads
\begin{equation}
    \hat{\eta} =-\frac{\beta_1\mathcal{P}}{\left(\beta_1 - \beta_2\right) \langle \dot{Q}_2\rangle}=-\frac{L_{21}f_2f_1+L_{22}f_2^2}{L_{11}f_1^2+L_{12}f_1f_2},
    \label{efff}
\end{equation}
from which $\eta=\hat{\eta}~\eta_c$ follows immediately. As previously, heat engine (${\cal P}<0$) and pump (${\cal P}>0$) regimes impose boundaries to the optimization with respect to $f_2$, whose absolute value must lie in the interval $0 \le \vert f_2 \vert \le \vert f_m \vert$, where $f_m = -L_{21}f_1/L_{22}$, i.e., the so-called stopping force for which ${\cal P}=0$. As previously, the optimization can be performed to obtain maximum power ${\cal P}_{\maxp}$ (with efficiency ${\eta}_{\maxp}$) or maximum efficiency ${\eta}_{\maxe}$   (with power ${\cal P}_{\maxe}$), by changing the force $f_2$  to optimal values $f_{2,\maxp}$ and $f_{2,\maxe}$, respectively. These optimal output forces can be expressed in terms of the Onsager coefficients as
\begin{equation}
  f_{2,\maxe}=\frac{L_{11} }{L_{12} }\left(-1+ \sqrt{1-\frac{L_{12}^2 }{L_{11} L_{22}}}\right)f_1,
  \label{eq:x2meta}
\end{equation}
and
\begin{equation}
  f_{2,\maxp}=-\frac{1}{2}\frac{L_{12}}{L_{22}}f_1,
  \label{x2mp}
\end{equation}
respectively, where the property $L_{21}=L_{12}$ has been considered. By inserting $f_{2,\maxe}$ or  $f_{2,\maxp}$ into the expression for $\hat{\eta}$, we obtain $\hat{\eta}_{\maxe}$ and the efficiency at maximum power $\hat{\eta}_{\maxp}$  given by
\begin{equation}
  \hat{\eta}_{\maxe}=-1+\frac{2L_{11}L_{22}}{L_{12}^2}\left(1-\sqrt{1-\frac{L_{12}^2}{L_{11} L_{22}}}\right),
  \label{etame}
\end{equation}
and
\begin{equation}
  \hat{\eta}_{\maxp}=\frac{L_{12}^2}{4 L_{11}L_{22}-2L_{12}^2},
\label{etamp}
\end{equation}
Similarly, we can derive the expressions for ${\cal P}_{\maxp}$ and ${\cal P}_{\maxe}$. All these quantities are not independent of each other, instead they satisfy the following relationships:
\begin{equation}
\hat{\eta}_{\maxp}=\frac{\hat{\eta}_{\maxe}}{1+\hat{\eta}^2_{\maxe}} \qquad {\rm and} \qquad \frac{{\cal P}_{\maxe}}{{\cal P}_{\maxp}}=1-\hat{\eta}^2_{\maxe},
\label{opti}
\end{equation}
where the symmetry between crossed Onsager coefficients $L_{12}=L_{21}$ has been taken into account. It is convenient to introduce the \textit{coupling parameter} $\kappa=L_{12}/\sqrt{L_{11}L_{22}}$ \cite{cleuren2015universality,TF9656101897}, in such a way that optimal efficiencies $\hat{\eta}_{\maxp}$ and  $\hat{\eta}_{\maxe}$ are solely expressed in terms of this quantity as follows
\begin{equation}
  \hat{\eta}_{\maxe}=-1+\frac{2}{\kappa^2}\left(1-\sqrt{1-\kappa^2}\right),
  \label{etame1}
\end{equation}
and
\begin{equation}
    \hat{\eta}_{\maxp} = \frac{1}{2}\frac{\kappa^2}{2-\kappa^2},
    \label{etamp1}
\end{equation}
respectively. Since $\langle {\dot \sigma} \rangle \ge 0$, it follows that $\kappa$ must be constrained in the interval $-1 \le \kappa \le 1$, implying that both $\hat{\eta}_{\maxp}$ and $\hat{\eta}_{\maxe}$ are confined to $0 \le \hat{\eta}_{\maxp} \le 1/2$ and $0\le\hat{\eta}_{\maxe}\le 1$, respectively. Notice that $\kappa = \pm 1$ implies that the determinant of the ($2 \times 2$) Onsager Matrix is equal to zero. This, in turn, implies proportionality between the two thermodynamic fluxes, i.e., $J_1 \propto J_2$, for all forces $f_1$ and $f_2$. Fig.~\ref{ons} shows that all the signatures about collective effects are also captured by the linear regime, describing very well the system behavior near the equilibrium regime (panels (b) and (c)). Remarkably, the increase of efficiencies towards the Carnot bound as $\epsilon$ and $F$ increase, as described in the main text, is understood from the interplay among Onsager coefficients, $L_{ij}$, that leads to $\kappa \rightarrow -1$ (panel (a) and inset). Also, ${\hat \eta}_{ME}$ and ${\hat \eta}_{MP}$ closely follow the analytical expressions presented in Eqs.~\eqref{etame1} and \eqref{etamp1} (see Fig.~\ref{ons}d).

\section{Crossover from heat engine to pump regimes}\label{pump} As described in the main text, the system operates as a pump 
when units operate almost (or completely) independently, or  (see Fig.~\ref{fig1}) as $F$ is raised. As $\epsilon$ increases towards positive values, the system hits a  threshold $\epsilon_c$ 
giving rise to the independent mode operation. It can emerge in different ways, such as via a discontinuous phase transition ($q=2$, model A and B; $q=3$, model B) or a continuous one ($q=3$, model A, $\alpha=1$), or even as a crossover with no phase transitions ($q=3$, model A, $\alpha \neq 1$). Although these results are exact in all cases, in the presence of a phase transition it is possible to obtain closed expression for ${\cal P}$ and $\langle \dot{Q}_\nu\rangle$ per unit. As shown in Sec. \ref{apb}, the disordered 
regime in such cases is characterized by 
by equal probabilities $p_{\downarrow}^*=p_0^*=p_\uparrow^*=1/3$ for $\epsilon \ge\epsilon_c$. By inserting into Eq. (\ref{work1}), it follows that
\begin{eqnarray}
\label{work2}
{\cal P} &=& 2 F \Big[\sinh(\frac{F \beta_1}{2}) + \sinh(\frac{F\beta_2}{2})\Big] \quad {\rm and} \nonumber \\
\langle \dot{Q}_\nu\rangle &=& -2 F \sinh(\frac{F \beta_\nu}{2}) \;,
\end{eqnarray}
respectively, both being independent on $\epsilon$. Similar formulas can be obtained for $q=2$ and $\epsilon \rightarrow \infty$, solely differing from them by a factor 2. The corresponding efficiency, in both cases, is $\eta=1+(\sinh(\beta_1 F/2)/\sinh(\beta_2 F/2))^{-1}$. All these expressions state that only a pump regime is possible when units operate independently.
Although both collective and independent operations allow the emergence of a pump regime, power and heat fluxes are independent from $\epsilon$ when units operate independently, Eq.~\eqref{work2}, indicating that, in the collective phase, $\epsilon$ can be chosen appropriately to lead to a better performance even as a pump. This result strenghten further the role of interactions and collective operations in an engine model with Ising-like interactions.

\section{Linear stability of disordered phase solution for models A and B  for q=3}\label{apb}
\begin{figure}[h]
\centering
\includegraphics[scale=0.3]{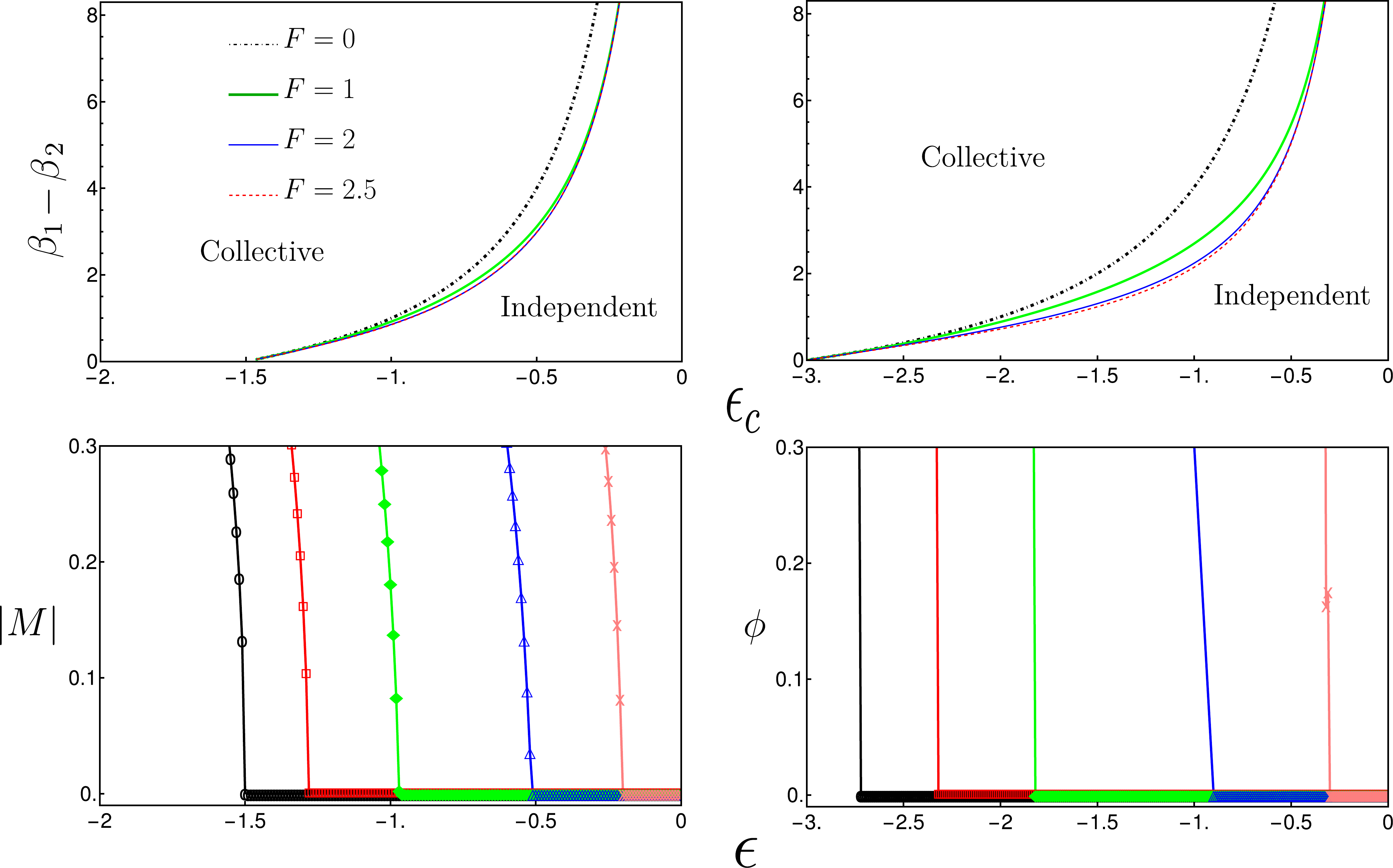}
\caption{For $q=3$ and  distinct $F$'s, left and right top panels show the phase diagrams $\Delta \beta=\beta_1-\beta_2$ versus $\epsilon_c$ for  model A and B, respectively. They are obtained from the linear stability analysis of the disordered phase. For the sake of comparison, the bottom panels show (for $F=1$) the  location of phase transitions from the order-parameter behaviors. From left to right $\beta_1-\beta_2 = 0$, $1/3$, $1$, $3$ and $9$. In all cases, we set $\beta_2=1$.}
\label{app1}
\end{figure}

For completeness, we provide additional information about the crossover between collective and independent regimes for model A (when $\alpha=1$) and B for $q=3$ which manifests through continuous and discontinuous phase transitions, respectively. These phenomena can be analyzed in a similar way to their equilibrium counterparts, by means of two order parameters, $M$ for model A and $\phi=({3}p_{max}-1)/2$ ($p_{max}={\rm max}\{p^{st}_\downarrow,p^{st}_0,p^{st}_\uparrow\}$) for model B, with the first one characterized by the classical exponent $\beta=1/2$ \cite{fiorejcp2013,challa}. However, contrasting to the equilibrium Potts model, nonequilibrium ingredients modify the phase transition for model B from a continuous to a discontinuous one, as shown in panel (b) of Fig.~\ref{app1}. 

A systematic investigation can be performed by means of a linear expansion of the master equation around a fixed point as follows ${\dot p}_m = \sum_n A_{mn}p_n$, where $A$ is the Jacobian matrix with elements $A_{mn} = \partial (\omega^{(1)}_{mn}+\omega^{(2)}_{mn})/\partial p_n|_{p_n=p^*}$ evaluated at fixed points $\sum_nA_{mn} p_n^*=0$. 
In particular, the solution $p_{n}^*$ is linearly stable if the real parts of the eigenvalues of the Jacobian matrix are negative. In both cases, the independent regime is characterized by equal population $p_{\downarrow}^*=p_0^*=p_\uparrow^*=1/3$ for $\epsilon \ge\epsilon_c$. In both  cases introduced above, the corresponding eigenvalues can be written as $\lambda_\pm=\lambda_0\pm \lambda_1$, with $\lambda_0$ given by
\begin{equation}
\lambda_0=-(3+\beta_1 \epsilon) \cosh \left(\frac{\beta_1 \
F}{2}\right)-(3+\beta_2 \epsilon) \cosh \left(\frac{\beta_2 F}{2}\right),
\end{equation}
whereas $\lambda_1$, for model A, reads:
\begin{align}
\lambda_1=\Big[&6+\epsilon ^2 \left(\beta_1^2+\beta_2^2\right)+\left(\beta_1^2 \epsilon ^2-3\right) \cosh
(\beta_1 F)-3 \cosh (\beta_2 F)+\beta_2 \epsilon ^2 \left(4 \beta_1 \cosh \left(\frac{\beta_1 F}{2}\right)
\cosh \left(\frac{\beta_2 F}{2}\right)+\beta_2 \cosh (\beta_2 F)\right)+\nonumber \\&+12 \sinh
\left(\frac{\beta_1 F}{2}\right) \sinh \left(\frac{\beta_2 F}{2}\right)\Big]^{1/2},
\end{align}
while, for model B, we have:
\begin{equation}
\lambda_1=i \sqrt{3}\left[\sinh \left(\frac{\beta_1 F}{2}\right)-\sinh \left(\frac{\beta_2 F}{2}\right)\right],
\end{equation}
Since $\lambda_1$ is imaginary for model B, the linear stability of disordered solution is granted provided $\lambda_0 < 0$. Conversely, for model A, due to the fact that $\beta_1$ and $\beta_2$ are always positive, $\lambda_-$ is always negative. Conversely, $\lambda_+$ is always negative for sufficiently large and positive $\epsilon$, with the order-disorder phase transition corresponding to a trans-critical bifurcation when $\lambda_+=0$. Clearly, $\lambda_+$ becomes positive as $\epsilon$ decreses, meaning that the independent regimes turns unstable.
    
Fig. \ref{app1} depicts the phase diagrams $\Delta\beta=\beta_1-\beta_2$ versus $\epsilon$ for different $F$ obtained from the linear analysis. In particular, for $F=0$, $\lambda_+$'s and $\lambda_-$'s read $-6$ and $-2[3+\epsilon(\beta_1+\beta_2)]$ (model A) and $-[6+\epsilon(\beta_1+\beta_2)]$ (model B), respectively, consistent to phase transitions taking place at $\epsilon_c=-3/(\beta_1+\beta_2)$ and $\epsilon_c=-6/(\beta_1+\beta_2)$. The crossover from collective to independent regime smoothly changes with the driving and it is more sensitive to the difference of temperatures. Note the excellent agreement between the values of $\epsilon_c$ obtained from the linear analysis and those from order parameter behaviors (bottom panels for $F=1$).

\section{Beyond the all-to-all case}\label{aph}
As described in the main text, the all-to-all case describes very accurately nearest-neighbor interactions in the regime of strong collective effects. We restrict, for simplicity, our analysis to model A and $q=3$ in a square lattice of linear size $L$. Each site $i$ is associated with a spin variable $\sigma_i=\pm 1, 0$. Then, Eq.~(1) of the main text becomes 
\begin{equation}
E_{i}=\frac{1}{2k}\sum_{i=1}^N\sum_{j=1}^k\epsilon\sigma_i\sigma_{i+j}\Big[ \delta_{\sigma_i,\sigma_{i+j}}+\alpha \delta_{\sigma_i,-\sigma_{i+j}}\Big].
\end{equation} 

\begin{figure}[H]
    \centering
    \includegraphics[scale=0.18]{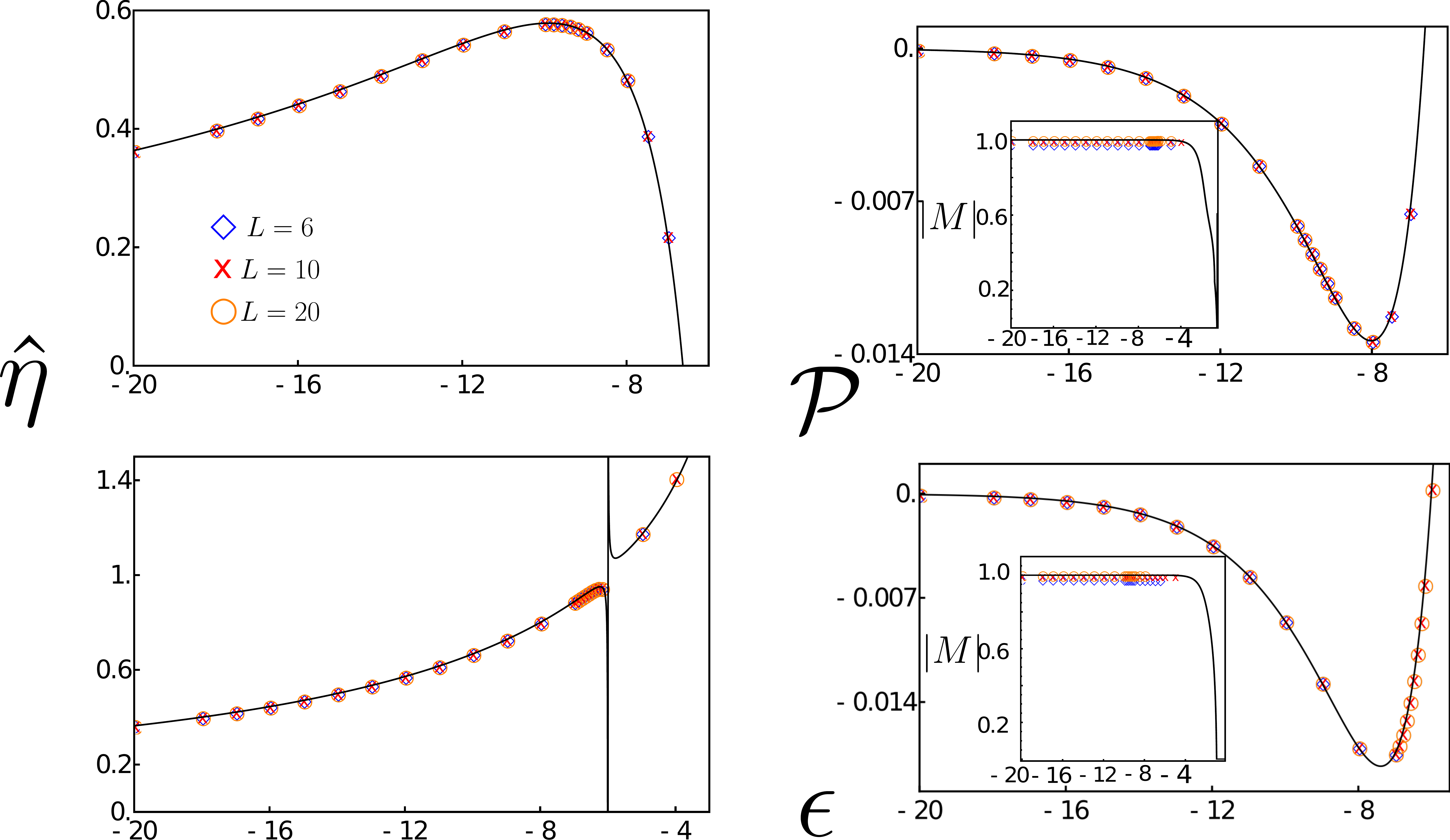}
    \caption{Performance of model A for a square lattice ($k=4$ nearest neighbors). Left and right panels depict efficiency ${\hat \eta}$ and power output ${\cal P}$ for distinct $\alpha = 1$ (top panels) and $\alpha = 3$ (bottom panels). Insets: The order parameter $|M|$ is shown. Symbols indicate numerical results for the square lattice of linear size $L$, i.e., $N = L^2$, while continuous lines the all-to-all case. Parameters: $\beta_1=2,\beta_2=1,F=2$. Numerical results were obtained from the Gillespie algorithm.}
    \label{figh}
\end{figure}

Despite the absence of exact results in such case, system's behavior and thermodynamic properties can be evaluated numerically by employing the Gillespie algorithm \cite{gillespie1977exact}. In Fig.~\ref{figh}, we compare these results with those in Fig.~1 of the main text for $F=2$, with $\alpha = 1$ (top panels) and $\alpha = 3$ (bottom panels), for increasing lattice size $L$. They agree almost perfectly, highlighting that the all-to-all case is insightful also when considering lattice models.

\end{document}